\documentclass[twocolumn]{aastex631}
\usepackage{amsmath,amssymb}
\usepackage{natbib}
\usepackage{graphicx}
\usepackage[figuresright]{rotating}
\usepackage{threeparttable}
\usepackage{multirow}
\newcommand{\OIII}{[O\,{\sc iii}]}

\newcommand{\Msun}{$M_{\odot}$}
\newcommand{\Zsun}{$Z_{\odot}$}

\newcommand{\Mh}{$M_\mathrm{h}$}
\newcommand{\HII}{H\,{\sc ii}}
\newcommand{\HI}{H\,{\sc i}}
\newcommand{\HH}{$\mathrm{H_2}$}

\newcommand{\fISM}{$f_\mathrm{ISM}$}
\newcommand{\fHI}{$f_\mathrm{HI}$}
\newcommand{\fHH}{$f_\mathrm{H2}$}
\newcommand{\UM}{{\sc UniverseMachine}}
\newcommand{\NUM}{{\sc NeutralUniverseMachine}}

\newcommand{\NAOJ}{National Astronomical Observatory of Japan, 2-21-1 Osawa, Mitaka, Tokyo 181-8588, Japan}
\newcommand{\SOKENDAI}{Department of Astronomical Science, The Graduate University for Advanced Studies, SOKENDAI, 2-21-1 Osawa, Mitaka, Tokyo 181-8588, Japan}
\newcommand{\UA}{Department of Astronomy and Steward Observatory, University of Arizona, Tucson, AZ 85721, USA}
\newcommand{\ICRR}{Institute for Cosmic Ray Research, The University of Tokyo, 5-1-5 Kashiwanoha, Kashiwa, Chiba 277-8582, Japan}

\accepted{on Mar 18, 2025}
\submitjournal{ApJ}
\shorttitle{ChemicalUniverseMachine I}
\shortauthors{Nishigaki et al.}
\graphicspath{{./}{figures/}}

\begin{document}

\title{ChemicalUniverseMachine I:\\
Uncovering the Cosmic Evolution of Metals in the Galaxy-ISM-CGM Ecosystem
}

\author[0000-0003-4321-0975]{Moka Nishigaki}
\affiliation{\NAOJ}\affiliation{\SOKENDAI}
\author[0000-0002-2517-6446]{Peter Behroozi}
\affiliation{\NAOJ}\affiliation{\UA}
\author[0000-0002-1049-6658]{Masami Ouchi}
\affiliation{\NAOJ}\affiliation{\ICRR}
\affiliation{Astronomical Science Program, Graduate Institute for Advanced Studies, SOKENDAI, 2-21-1 Osawa, Mitaka, Tokyo 181-8588, Japan}
\affiliation{Kavli Institute for the Physics and Mathematics of the Universe (Kavli IPMU, WPI), The University of Tokyo, 5-1-5 Kashiwanoha, Kashiwa, Chiba, 277-8583, Japan}
\author[0000-0003-4936-8247]{Hong Guo}
\affiliation{Shanghai Astronomical Observatory, Chinese Academy of Sciences, Shanghai 200030, People's Republic of China}
\author[0000-0002-6748-6821]{Rachel S. Somerville}
\affiliation{Center for Computational Astrophysics, Flatiron Institute, 162 5th Avenue, New York, NY 10010, USA}
\author[0000-0002-9656-1800]{Anna R. Gallazzi}
\affiliation{INAF-Osservatorio Astrofisico di Arcetri, Largo Enrico Fermi 5, I-50125 Firenze, Italy}
\author[0000-0003-2965-5070]{Kimihiko Nakajima}
\affiliation{\NAOJ}
\affiliation{Institute of Liberal Arts and Science, Kanazawa University, Kakuma-machi, Kanazawa, Ishikawa, 920-1192, Japan}
\author[0000-0002-2740-3403]{Kuria Watanabe}
\affiliation{\NAOJ}\affiliation{\SOKENDAI}

\begin{abstract}
We present an empirical chemical evolution model that explains the distribution of metals in the interstellar medium (ISM) and the circumgalactic medium (CGM) of galaxies based on the {\sc UniverseMachine} and {\sc NeutralUniverseMachine} models in the framework of $\Lambda$CDM structure formation. We parameterize the fractions of outflowing metals returned and mixed into the multi-phase ISM of the star-forming regions ($f_{\rm H2}$) and into the neutral gas regions ($f_{\rm HI}$); metal production, transfer, and dilution are caused by star formation, galaxy mergers, and gas inflow from the inter-galactic medium, respectively, with rates determined by the {\sc (Neutral)UniverseMachine} models.
Using a Markov Chain Monte Carlo algorithm, we explore the posterior distributions of metal return and mixing consistent with observed mass-metallicity relations in H\,{\sc ii} regions (at $0<z<5$), H\,{\sc i} damped Lyman-alpha systems (at $1<z<4$), and the CGM (at $z=0$). 
We find that the fraction of metals present in the ISM, $f_{\rm H2}+f_{\rm HI}$, increases with halo mass from $\sim20$\% at $10^{10}M_\odot$ to $\sim80$\% at $10^{13}M_\odot$. These fractions increase mildly at higher redshifts, to $\sim30$\% at $10^{10}M_\odot$ and $\sim80$\% at $10^{13}M_\odot$ at $z=5$. Interestingly, there is no significant redshift evolution of $f_{\rm H2}+f_{\rm HI}$ at fixed circular velocity, suggesting that metal distribution between the ISM and CGM is universally determined by the halo potential well depth.
CGM metal enrichment is thus slow in high-$z$ halos with deep potential wells. While $f_{\rm H2}$ monotonically increases with halo mass, $f_{\rm HI}$  peaks at $\sim10^{12}-10^{13} M_\odot$, suggesting that reinfall may be inefficient in larger-mass halos.
\end{abstract}


\section{Introduction} \label{sec:intro}
Chemical evolution of the interstellar medium (ISM) and circumgalactic medium (CGM) provides crucial insights into physical processes governing galaxy formation and evolution. Metallicities in the ISM and CGM are influenced by processes such as star formation, metal loss by outflows from the ISM to the CGM, and metal dilution by inflows from the inter-galactic medium (IGM) or CGM.  Hence, they have long interested astronomers (see \citealt{Maiolino19} for a review).

Metallicity observations have been actively advancing, now covering many different regions of galaxies and extending from the distant universe to the present day. Metallicities in the \HII\ regions, determined from emission line ratios in optical spectra, have been most well studied in galaxies from the Sloan Digital Sky Survey (SDSS) at $z\sim0$. It has been found that there is a strong scaling relation between mass and metallicity, known as the mass-metallicity relation \citep[MZR,][]{Tremonti04,Mannucci10,AM13,Curti20}. Ground-based telescopes have examined the metallicity of galaxies up to $z\sim3$ \citep{Erb06,Maiolino08,Nakajima13,Ly16,Sanders20,Sanders21}. More recently, thanks to the James Webb Space Telescope (JWST), it has become possible to investigate the metallicity of galaxies beyond $z\sim4$ \citep{Curti23,Curti24,Fujimoto23,Nakajima23,Sanders24}. These results indicate that the MZR evolves with redshift, where metallicity decreases as redshift increases for a given mass.
In \HI\ regions, metallicities are derived from the absorption lines of Damped Lyman-Alpha (DLA) systems \citep{Wolfe05,Ledoux06,Banados19,Berg21}. It is known that a MZR exists in \HI\ regions up to $z\sim6$ \citep{Ledoux06,Moller13}.
In the CGM, metallicities are also determined from absorption lines \citep[e.g.,][]{Tumlinson17,Prochaska17}. However, since identifying a host galaxy is challenging, there are limited studies on the relationship between the mass of host galaxies and the metallicities of the CGM. Metallicity measurements in multi-phase gas of galaxies provide valuable information for constraining galaxy evolution processes.

However, observational data offer only snapshots of these processes, whereas we would like to understand how gas flows evolve over time. 
{
There are various approaches to model the chemical evolution of galaxies, including analytical models \citep[e.g.,][]{Dave12,Lilly13}, semi-analytical models \citep[e.g.,][]{Somerville08}, and hydrodynamical simulations \citep[e.g.,][]{Finlator08,Ma16,Torrey19}. These models have strong physical priors regarding star formation and feedback, which have made it difficult to interpret both agreements and tensions with observations.
}
As such, the framework of empirical modeling is powerful tool. These models are constructed to infer the underlying relationships between dark matter structure formation and observed properties of galaxies. Because structure formation can be evolved accurately over time, these empirical relationships provide a unique window into how populations of galaxies (and their gas) evolve over time. Unlike semi-analytic models and hydrodynamical simulations, empirical models do not have strong physical priors and search flexible parameter spaces to infer the relationships present in the real Universe.   

The \UM{} \citep{Behroozi19} is a recent example. The model starts with dark matter halo merger trees generated from N-body simulations, which provide the formation histories of dark matter halos. The \UM\ employs a flexible approach to assign star formation rates (SFRs) to galaxies based on their host halo properties and their formation histories. To constrain the relations between dark matter halos and galaxies, the \UM\ is calibrated using a wide array of observational data, including galaxy stellar mass functions, star formation rates, quenched fractions, and galaxy clustering across a wide range of redshifts.

Building upon the framework of the \UM, \cite{Guo23} present the \NUM, an empirical model that focuses on the properties and evolution of neutral and molecular hydrogen within galaxies across cosmic time. Similar to the \UM, the \NUM\ employs a flexible approach to assign \HI\ and \HH\ masses to galaxies based on their host halo properties, formation histories, SFRs, and stellar masses. The \NUM\ is calibrated using a comprehensive set of observational data, including \HI\ and \HH\ mass functions, the molecular-to-atomic ratio, the \HI–halo mass relation, \HI/\HH–stellar mass relations at $z \sim 0$, as well as the evolution of cosmic gas densities $\rho_\mathrm{HI}$ and $\rho_\mathrm{H2}$ at $0 < z < 6$.
Together, the \UM\ and \NUM\ can make robust inferences for the stellar and gas properties of galaxies over cosmic time.

This paper presents a model of metallicity evolution in the multiphase ISM and the CGM of galaxies at $z=$ 0--5 and discusses the metal distribution within these regions. Specifically, we construct a model for chemical evolution of galaxies by adding chemical evolution processes to the \UM\ and \NUM\ models. 
{
We parameterize the fractions of metals distributed and mixed into the multi-phase ISM of the star-forming regions ($f_{\rm H2}$) and the neutral gas regions ($f_{\rm HI}$), with a fraction $1-(f_{\rm H2}+f_{\rm HI})$ ejected directly from the galaxy into the CGM. In this paper, we refer to \HH{} as the gas in a galaxy that is actively forming stars, while \HI{} represents other gas in the ISM that may be physically separate and chemically distinct.
}
Metal production, supply, and dilution are caused by star formation, galaxy mergers, and gas inflow from the inter-galactic medium, respectively, with rates determined by the {\sc (Neutral)UniverseMachine} models. Since the amount of gas at each redshift is already constrained empirically by the \NUM\ model, net gas inflow and outflow rates are already constrained by the evolution of structure in $\Lambda$CDM.

{
There are two different treatments of mass and metal outflows discussed in the literature on galactic chemical evolution and the mass-metallicity relation. One approach assumes that galactic winds eject the ISM with a mass-loading factor $\eta$, resulting in a metal ejection rate of $\eta Z_\mathrm{out} \mathrm{SFR}$, where $Z_\mathrm{out}$ is metallicity of outflow gas \citep[e.g.,][]{Finlator08,Peeples11,Dave12}. The alternative approach assumes that a fraction of supernova yields is lost directly, leading to a decoupling between the metal-loss rate and the mass outflow rate \citep[e.g.,][]{Schonrich09}.
In this work, we adopt the latter approach, which allows us to describe metal ejection in terms of $f_{\rm H2}$ and $f_{\rm HI}$. This enables a more direct treatment of metal loss without requiring assumptions or parameterizations about gas regulation physics, such as mass-loading factors.
}

In this first paper in the \textsc{ChemicalUniverseMachine} series, we seek to empirically infer the average behavior of galaxy populations as a function of host halo mass and redshift; further papers will use constraints on galaxies' average behavior to inform the parameterization for how individual galaxies evolve.

We describe our methodology in Section \ref{sec:method}. Section \ref{sec:data} presents the observational data that we use. Our main results are shown in Section \ref{sec:result}, followed by discussion in Section \ref{sec:discussion}. We summarize our conclusions in Section \ref{sec:summary}. Throughout this paper, we assume a \cite{Chabrier03} initial mass function (IMF). We adopt a flat CDM cosmology with $\Omega_m = 0.307, \Omega_\Lambda = 0.693, h = 0.678, \sigma_8 = 0.823, n_s = 0.96$, consistent with Planck results \citep{Planck15}. 
Halo masses are defined based on the \cite{Bryan98} spherical overdensity definition and denote peak historical halo masses extracted from the merger tree ($M_\mathrm{peak}$).
We use the solar abundance ratios of \cite{Asplund09}.
\section{Method} \label{sec:method}
\subsection{Method Overview}
\label{sec:method_overview}
\begin{figure}[ht]
    \centering
    \includegraphics[width=\linewidth]
        {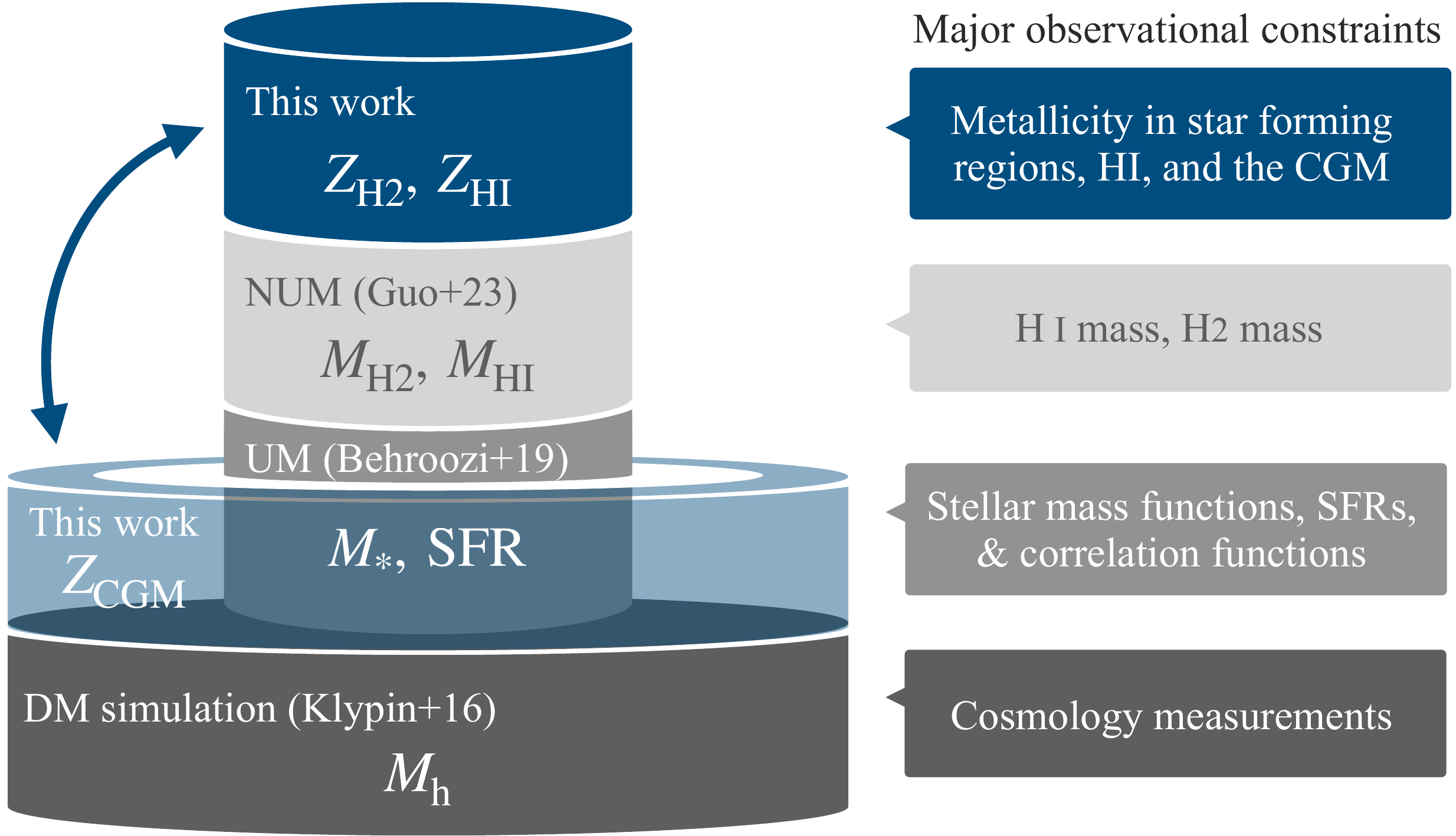}
    \caption{
    Schematic representation of model hierarchical structure. The metallicities in the ISM (i.e., \HI\ and \HH\ regions) and CGM are constrained in this work. Dark matter halo properties, such as halo masses and assembly, are provided from the dark matter simulation \citep{Klypin16,Rodrguez-Puebla16b}. Stellar properties, including star formation rates (SFRs) and stellar masses, are obtained from the \UM\ \citep{Behroozi19}. For the ISM mass and its time evolution, we use the constraints from the \NUM\ \citep{Guo23}. The major observational constraints for each model are shown on the right side.
    }
    \label{fig:schem}
\end{figure}

Our model (Fig.\ \ref{fig:schem}) constrains the evolution of average ISM and CGM metallicities as a function of halo mass and redshift.  To understand why we can constrain the metal budget of galaxies empirically, it is helpful to consider an approximation to the full differential equations. The total mass of metals produced in a galaxy can be approximately determined as the product of the stellar mass ($M_*$) and the yield ($y$), where the yield here represents the fractional mass of metals injected into the ISM per unit mass of stars formed. Defining the parameter \fISM\ to be the fraction of metals returned and mixed into the ISM (as opposed to ejected into the CGM or beyond), we can approximately relate the ISM metallicity ($Z_\mathrm{ISM}$) to the gas mass ($M_\mathrm{ISM}$) as 
\begin{equation}
Z_\mathrm{ISM} \sim \frac{\langle f_\mathrm{ISM}\rangle \ y\ M_*}{M_\mathrm{ISM}}. \label{eq:basic}
\end{equation}
Hence, given measurements of a galaxy's stellar mass, total ISM gas mass, and gas-phase metallicity, we could estimate the average metal distribution fraction over its lifetime, without the need to build an empirical model. If we had such measurements for a range of galaxies, such that we cover a range of halo masses and redshifts, there is then sufficient information to determine how the ISM distribution fraction varies with halo mass and redshift.

In this paper, we solve the full metallicity evolution equation instead of the approximation above, accounting for mergers and inflowing gas. However, our constraints come from the same basic observational information. In detail, stellar and ISM gas properties, including stellar and ISM gas masses as functions of halo mass and redshift (i.e., $M_*(M_\mathrm{h}, z)$ and $M_\mathrm{ISM}(M_\mathrm{h}, z)$), are well-constrained by previous models. We utilize the best-fitting \UM\  \citep{Behroozi19} model for galaxy stellar properties (Section \ref{sec:UM}) and the best-fitting \NUM\ \citep{Guo23} model for ISM gas properties (Section \ref{sec:NUM}), both calculated on dark matter halos in a dark matter simulation (Section \ref{sec:DMsimu}). These models are briefly described in the following three sections. 
Anchored by these models, we constrain ISM metallicity as a function of halo mass and redshift (i.e., $Z_\mathrm{ISM}(M_\mathrm{h}, z)$) with observational constraints on ISM and CGM metallicity, free parameters for \fISM, and an assumption for the yield $y$. In our models, a fraction $1-f_\mathrm{ISM}$ of the metal mass is ejected into the CGM. In other words, \fISM\ quantifies the proportion of metals distributed between the ISM and CGM. Table \ref{tab:models} summarizes observational constraints and redshifts of both previous and our models.

\subsection{Multi-phase ISM Overview}

The most common approach to measure ISM metallicity is to derive metallicity from emissions of ionized gas in star-forming regions. However, these measurements may not fully represent the metallicity in the entire ISM, as the gas exists in other phases, including molecular and atomic. It is believed that stars form from molecular gas, and as stars within molecular clouds evolve, they emit ionizing radiation, creating \HII\ regions. Consequently, we assume that both ionized gas and molecular gas coexist in star-forming regions and that metals are well-mixed in these regions. On the other hand, neutral gas is believed to be stored in the surrounding region, as suggested by observations \citep{Walter08,Neeleman17} and simulations \citep{Bird14,Rahmati14}. The mixing of metals within star-forming regions (\HII\ and \HH) with \HI\ regions and the CGM remains not well understood.

In this paper, we present two models: one adopts a simpler assumption that metals are well-mixed within the entire ISM (Section \ref{sec:wellmix_model}), and another employs a more realistic approach where metals in \HH\ and \HI\ regions are treated separately (Section \ref{sec:realistic_model}). For both models, we use observational constraints on CGM metallicity, as well as those on ISM metallicity, to gain insights into the mixing of metals between the ISM and CGM. In the latter model, we include constraints from metallicity measurements in \HI\ regions, offering empirical insight into metal mixing within the ISM.

We refer to the ISM mass as the total mass of molecular and atomic gasses, while the mass of ionized gas is ignored as it is negligibly small compared to the total ISM mass \citep{McKee77}. The CGM is referred to as the region outside of the ISM and inside of the halo virial radius, and the mass of the CGM is determined by the mass difference between the total baryonic halo mass (estimated as the product of the halo mass and the cosmic baryon fraction) and the stellar$+$ISM mass. We refer to the IGM as the regions outside of the halo virial radius.
\begin{table*}[ht]
    \centering
    \begin{tabular}{llcl}
        \hline\hline
        Name & Model Outputs & Major Observational Constraints & Redshift Range \\\hline
        {\it Bolshoi–Planck} & DM halo properties & Cosmology measurements & 0--10 \\
        \UM\ & SM \& SFR & SMFs, 2PCFs, \& SFRs & 0--10\\
        \NUM\ & \HI\ \& \HH\ masses & \HI\ \& \HH\ measurements & 0--6\\
        \textsc{ChemicalUniverseMachine} & ISM \& CGM metallicities & Metallicity in star-forming regions & 0--6\\\hline
    \end{tabular}
    \begin{tablenotes}
    \item {Notes.} DM represents dark matter, SM refers to stellar mass, SMF stands for the stellar mass function, and 2PCF is the abbreviation for the two-point correlation function.
    \end{tablenotes}
    \caption{Summary of theoretical models.}
    \label{tab:models}
\end{table*}
\subsection{Dark Matter Simulation}
\label{sec:DMsimu}
The \UM\ relies on the {\it Bolshoi–Planck} simulation \citep{Klypin16,Rodrguez-Puebla16b} for halo properties and assembly histories. Covering a periodic, comoving volume of 250 $h^{-1}$ Mpc on each side with $2048^3$ particles ($\sim8\times10^9$), this simulation, conducted using the {\sc ART} code \citep{Kravtsov97,Kravtsov99}, has high resolutions in mass ($1.6 \times 10^8 h^{-1} $\Msun), force (1 $h^{-1}$ kpc), and time output (180 snapshots equally spaced in $\log(a)$). The adopted cosmology is flat $\Lambda$CDM with parameters of $h = 0.678$, $\Omega_m = 0.307$, $\sigma_8 = 0.823$, and $n_s = 0.96$, aligning with the {\it Planck}15 results \citep{Planck15}. 
The {\sc rockstar} \citep{Behroozi13b} and {\sc Consistent Trees} \citep{Behroozi13d} codes are used for halo finding and merger tree construction, respectively.
\subsection{UniverseMachine}
\label{sec:UM}
The \UM\ \citep{Behroozi19} is an empirical framework that establishes the connection between galaxy growth and halo growth with observational constraints spanning a redshift range of $z=$ 0--10. This model parameterizes galaxy SFRs as a function of halo potential well depth, redshift, and assembly history for individual halos in the dark matter simulation. The galaxy stellar mass is self-consistently derived from the star formation history along a halo’s assembly and merger history for each individual halo. The parameter space posterior distribution is determined using a MCMC method by predicting the observable SFRs and stellar mass abundances for a given point in parameter space and then comparing with real observations across a wide range of redshifts. The major observational constraints for the \UM\ include stellar mass functions, specific SFRs, and galaxy auto- and cross-correlation functions. 

The \UM\ output is complied as a catalog,\footnote{\url{https://halos.as.arizona.edu/UniverseMachine/DR1}} providing halo and galaxy properties for individual halos at each timestep, including halo mass, SFR, and stellar mass. We extract the average SFR and stellar mass at a given halo mass and redshift. For the definition of halo masses, we use peak historical virial masses extracted from the merger tree ($M_{\mathrm{peak}}$).
\subsection{NeutralUniverseMachine}
\label{sec:NUM}
\begin{figure}[t]
    \centering
    \includegraphics[width=\linewidth]
        {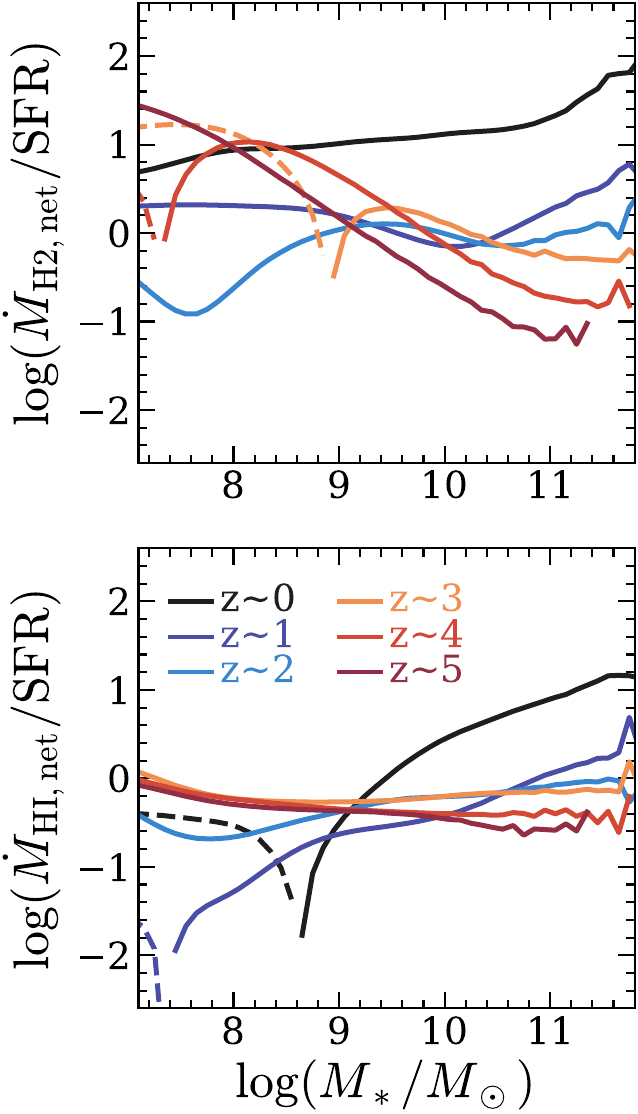}
    \caption{{Net gas inflow/outflow rates divided by SFR as functions of stellar mass and redshift, averaged for star-forming galaxies. Solid lines represent cases where inflow exceeds outflow, while dashed lines indicate the opposite.}}
    \label{fig:num}
\end{figure}

The \NUM\ \citep{Guo23} is an empirical model framework that constrains the evolution of atomic hydrogen gas (\HI) and molecular hydrogen gas (\HH) masses, developed on the halos in the \UM\ catalog. 
{
The net gas inflow rates are governed by the evolution of structure in $\Lambda$CDM.
Figure \ref{fig:num} presents the ratios of net gas inflow rates to SFRs as functions of stellar mass and redshift. We calculate the net inflow rates by taking the difference in average gas mass for star-forming galaxies between consecutive timesteps and dividing by the time interval between them (i.e., $\dot{M}_\mathrm{gas,net}=(M_\mathrm{gas,in} - M_\mathrm{gas,out}) / \Delta T$). 
}

The \HI\ mass of each halo is parameterized as a function of halo mass, SFR, redshift, and halo formation history. The \HH\ mass is parameterized as a function of stellar mass, offset from the star formation main sequence, and redshift. The parameter space posterior distributions are determined by comparing the  implied gas masses and their abundances to observations across a wide range of redshifts. The \NUM{} model uses observational constraints from various sources, including the \HI\ and \HH\ mass functions, the molecular-to-atomic ratio, the \HI–halo mass relation, and the \HI/\HH–stellar mass relations at $z \sim 0$. This model also incorporates the evolution of cosmic gas densities $\rho_\mathrm{HI}$ and $\rho_\mathrm{H_2}$ across the redshift range of $0<z<6$.  In practice, much of the gas-to-halo mass constraining power of the \NUM\ arises because it combines observed gas mass measurements (as functions of stellar mass, SFR, and redshift) with the stellar-to-halo mass relations from the \UM{}. 

Observations of \HI\ in the local universe come from the ALFALFA survey \citep{Haynes18}, while measurements of \HH\ in the local universe are originate from CO observations in the xCOLD GAS survey \citep{Saintonge17}. The cosmic density of \HI\ is determined through 21 cm measurements in the local universe and through measurements of damped Lyman-alpha systems (DLAs) at high redshifts. The cosmic density of \HH\ is inferred from CO and dust continuum measurements.

The \NUM\ catalog provides \HI\ and \HH\ masses for every halo in the \UM{} catalog. We obtain the average \HI\ and \HH\ masses at a given halo mass and redshift from the \NUM{} catalog. For both \HI\ and \HH\ masses, we consider masses of hydrogen without accounting for the contribution of helium and heavier elements, as  observed metallicities are expressed relative to the hydrogen content of the gas.
\subsection{Computing Metallicity Evolution}
\label{sec:method_evolution}
We constrain the average ISM and CGM metallicities in each halo mass and redshift bin for both star-forming galaxies (SFGs) and quenched galaxies (QGs). The calculations are based on the average SFRs and ISM masses, which are obtained from the \UM\ and \NUM\ catalogs, respectively. Galaxies are moved between SFG and QG categories based on their histories in the \UM. While only the metallicities of star-forming galaxies are compared to observational data, metallicities of quenched galaxies are also tracked to consider populations that become star forming again after being quenched (i.e., those that rejuvenate). We define star-forming galaxies as galaxies with $\mathrm{sSFR}>10^{-11}\ \mathrm{yr}^{-1}$.

We propagate the average values in each halo mass and redshift bin \citep[as in][]{Behroozi13e}, rather than calculating for individual halos as done in the \UM. As well, the constraints for SFR, $M_*$, $M_{\mathrm{HI}}$, and $M_{\mathrm{H_2}}$ from the \UM\ and \NUM\ are fixed, instead of re-fitting them simultaneously with the metallicity calculations. This approach is ideal for our goal of understanding the average evolution of metallicities, as it
enables $\sim 10,000\times$ faster model evaluation than computing individual galaxy properties. \cite{Behroozi13e} discuss that the impact of averaging different SFHs in bins of halo mass is minor compared to observational uncertainties. Since we only predict the average metallicities of galaxy populations using this approach, we use the observational uncertainties for the average metallicities instead of the galaxy-to-galaxy metallicity scatter when comparing to observations (see also Section \ref{sec:data}). As noted in the introduction, tracking chemical evolution for individual halos will be performed in future papers in this series.

The evolution of the gas-phase metallicity in the ISM arises from several physical processes. Here, we consider the effects of new star formation, metal loss by outflow, halo assembly, and accretion from the CGM. The metal mass evolution in the ISM is expressed as follows:
\begin{eqnarray}
    \label{eq:mz_ism}
    \dot M_{Z,\mathrm{ISM}}
    &=& 
    \{yf_\mathrm{ISM} - Z_\mathrm{ISM}(1-R)\} \mathrm{SFR}\\ \nonumber
    &+& Z_\mathrm{inf}\dot M_\mathrm{inf\rightarrow ISM}
    + Z_\mathrm{ISM,merger}\dot M_\mathrm{merger}.
\end{eqnarray}
In equation \ref{eq:mz_ism}, the metal mass produced by new star formation is determined by the galaxy's SFR and stellar yield $y$. The stellar yield refers to the fractional mass of metals injected into the ISM per unit mass of stars formed. 
$R$ denotes the return mass fraction, defined as the proportion of total mass that a stellar generation returns to the ISM.
We adopt constant values of $y = 0.016$ and $R=0.46$ for the stellar oxygen yield with the \cite{Chabrier03} IMF, following \cite{Vincenzo16}. We assume that the oxygen is promptly ejected into the ISM shortly after the beginning of star formation, because oxygen is mainly produced by short-lived massive stars undergoing core-collapse supernovae and the contributions from type Ia supernovae and AGB stars are negligible \citep{Peeples14}. 

Of the metal (i.e., oxygen) mass {ejected by supernovae}, a fraction \fISM\ distributes to the ISM, while the remaining fraction $1-$ \fISM\ is ejected into the CGM, with \fISM\ being a free parameter. Star formation also removes some small amount of metals from the ISM due to metals being locked up in stars.
{
The metals that are already present in the ISM from previous enrichment episodes and removed from the ISM by a wind correspond to a negative \fISM. However, since \fISM\ also accounts for metal reinfall and a more sustained distribution, it is expected to be positive. A negative \fISM\ would rapidly deplete metals from the ISM, whereas observed galaxy populations show an increasing metal mass over time. We further discuss this in Section \ref{sec:comparison}.
}
Lastly, $Z_\mathrm{inf}$ and $\dot{M}_\mathrm{inf}$ refer to the metallicity and accretion rate of infalling gas, which could come from the IGM, the CGM, or some combination of both.

As galaxies mature, the gas-phase metallicity reaches a limiting state (i.e., saturation). The saturation metallicity $Z_\mathrm{sat}$ is expressed as $Z_\mathrm{sat} = yf_\mathrm{ISM}/(1-R)$, assuming rapid mixing with the ISM during star formation. Once the ISM metallicity reaches the saturation level, it is difficult for it to increase much further, as the metals produced by star formation are exactly balanced by the metals locked up in new stars, and the only new sources of metals are mergers and accretion. 

\subsection{Accounting for Halo Growth and Mergers}
\label{sec:method_mergers}
\begin{figure}[t]
    \centering
    \includegraphics[width=\linewidth]
        {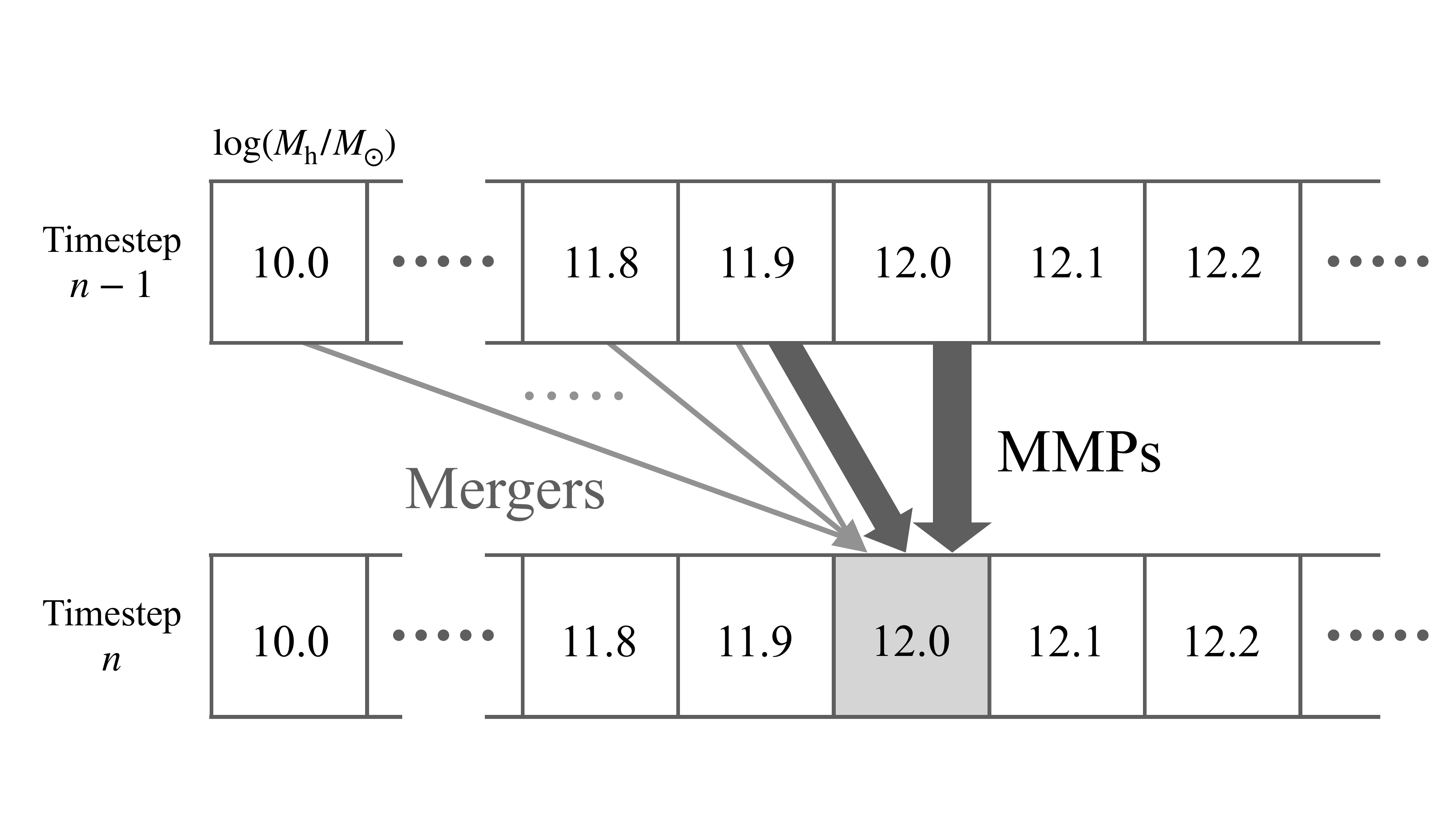}
    \caption{Schematic of gas and metal masses propagation in halo assembly histories.
    In this work, we average halo assembly histories in bins of halo mass, redshift, and star-formation status (star-forming or quiescent) to accelerate model evaluation. Gas and metal masses from most-massive progenitors are propagated to descendant halos at the next redshift, and the corresponding masses for merging satellites are distributed to the descendant halo's CGM or ISM depending on how far the satellites are from their host halos at the last snapshot before disruption (see text).}
    \label{fig:assembly}
\end{figure}

We calculate the metal mass accretion through halo assembly using the \UM\ catalog, which provides information on individual halos at each timestep (i.e., redshift) of the dark matter simulation, including their mass and progenitors. In each timestep, the average metal mass and ISM mass within each halo mass bin are carried over to the next timestep as halo assembly progresses. Figure \ref{fig:assembly} schematically illustrates how gas and metal masses are propagated to a given \Mh\ bin at timestep $n$. We average the gas and metal masses for the most massive progenitors (MMPs) for all halos in a given \Mh\ bin.

Mergers primarily impact massive halos with $M_h\gtrsim10^{13} M_\odot$; below this halo mass, mergers bring in only small fractions of the total stellar, gas, and metal mass \citep{Behroozi19}. Merging galaxies may either disrupt as stellar streams (contributing stellar mass to the intrahalo light) or merge with the central galaxy. In the \UM, merger outcomes are determined by the location where the merging satellite halo falls below the disruption threshold. If this distance $R$ is less than $0.4 R_\mathrm{vir}$ (where $R_\mathrm{vir}$ is the host halo's virial radius), the satellite merges into the central galaxy; otherwise it merges into the intrahalo light.  This threshold is constrained empirically from the clustering of galaxies (which determines the survival time of merging satellites) and the increase in massive galaxy number densities over time (which determines the fraciton of mergers that end up in the central galaxy vs.\ the intrahalo light). We adopt the same distance threshold for determining the outcome for gas and metal masses during satellite mergers.  It is worth noting that not all gas in mergers may cool to the ISM of the host galaxies, even when stars are disrupted. Confining this fraction requires additional observational constraints. Instead, we prepare models with two extreme conditions: one where the ISM of all satellites with $R < 0.4\ R_\mathrm{vir}$ is integrated into the host galaxy's ISM, and the other where all merging ISM is integrated into the host galaxy's CGM.
\subsection{Inflows and CGM Metallicity}
\label{sec:inflows_cgm}
We also incorporate metals present in inflowing gas from the CGM and IGM, although the {metallicities} in both gas reservoirs are typically small and thus have little impact on our model. Since observational constraints on the fraction of gas incoming from the IGM vs.\ the CGM are limited, we consider two extreme models where all inflowing gas originates from 1) the CGM, or 2) the IGM. To evaluate the amount of metals transported through inflows, we utilize the time differences in ISM mass. If the ISM mass obtained from the \NUM\ catalog exceeds that from progenitors, we infer that the net excess gas is supplied by inflows.\footnote{If there are additional inflows balanced by outflows from the galaxy, we assume that they do not contribute any extra metals, as it is difficult to remove gas without removing the associated metals as well.  The opposite physical situation of additional inflows mixing with ISM gas and being blown out---contributing to more efficient metal removal---is incorporated into the definition of $f_\mathrm{ISM}$.} We assume that the IGM metallicity is 0.01 \Zsun, as there is evidence that the IGM contains metals \cite[e.g.,][]{Simcoe04} and that therefore inflowing gas probably also contains metals. 
{
We show in \ref{sec:each_model} that setting inflow metallicity to $Z_\mathrm{CGM}$, 0.01, or zero makes little difference to the best-fit values of model parameters or to the quality of the fits.
}

The metals in the CGM originate from progenitors (i.e., merging satellites) and outflows from galactic star formation. We define the CGM mass as the difference between total baryonic halo mass and the sum of ISM and stellar masses (i.e., $M_\mathrm{CGM} = f_b M_\mathrm{h} - M_\mathrm{ISM} - M_*$, where $f_b = 0.15$ is the cosmic baryon fraction, and in this case we account for helium in the mass of the ISM). We note that galactic outflows may transport both gas mass and metal mass out of the CGM. In this paper, lacking strong observational constraints, we assume that outflows transport an equal fraction of metals and gas mass out of the CGM, so that the CGM metallicity is unchanged by beyond-CGM outflows.

Metal mass evolution in the CGM is expressed as:
\begin{eqnarray}
    \dot{M}_{Z,\mathrm{CGM}}
    &=& (1-f_\mathrm{ISM})y \mathrm{SFR} \nonumber\\
    &-& Z_\mathrm{CGM}\dot M_\mathrm{HI\rightarrow CGM}
    + Z_\mathrm{CGM,merger}\dot M_\mathrm{merger}.
    \label{eq:zcgm}
\end{eqnarray}
The metal mass in the CGM is propagated to the next timestep. This includes CGM metals of all progenitors, as well as ISM metals of progenitors that are disrupted outside of the host galaxy's ISM (i.e., at $R > 0.4\ R_\mathrm{vir}$). Additionally, if the ISM mass obtained from the \NUM\ catalog is decreasing with time in a given halo mass bin beyond the rate of star formation, the remaining excess balance of gas is assumed to be transported to the CGM, along with a metal mass corresponding to the ISM metallicity. Lastly, in one of our considered models, metals may also be lost to the CGM due to excess inflow into the ISM from mergers.

We found that rejuvenating galaxies require careful treatment due to our method of averaging galaxy properties. This is because, in the \UM, galaxies can undergo a slow quenching process where they oscillate between being star-forming and quenched for multiple Gyr before they quench completely. Since quenched galaxies have significantly lower average ISM masses than star-forming galaxies, the simplest mass binning scheme, 
where the inherently broad distribution of SFRs is condensed into a single average SFR (similar to a delta distribution),
would require substantial fresh low-metallicity accretion from the IGM/CGM every time a quenched galaxy returned (however briefly) to forming stars, resulting in unphysically low metallicities for massive star-forming galaxies. To address this issue, we assume that quenching galaxies lose gas in the central star-forming regions (shutting off star formation) but that the ejected gas and metals remain nearby. The gas and metals then return to the central region when the galaxies temporarily resume star formation.
\subsection{Well-Mixed ISM Model}
\label{sec:wellmix_model}
In this paper, we test two models for the distribution of metals in the ISM.  In the first model, which we call the ``Well-Mixed ISM Model,'' we assume that metals in the ISM, including ionized, molecular, and atomic hydrogen, are well-mixed. We calculate the ISM mass as the sum of the molecular and atomic hydrogen masses in the \NUM\ catalog.

In this model, the only free variable is the ISM distribution fraction \fISM, which we parameterize as a function of halo mass and redshift ($f_\mathrm{ISM}(M_\mathrm{h}, z)$). We have found that a double power-law function effectively reproduces the MZR at $z \sim 0$. Therefore, we adopt the following parameterization:
\begin{equation}
    f_\mathrm{ISM} = \frac{1}{1+(M_\mathrm{h}/C_0)^\alpha(1+z)^\beta}.
    \label{eq:fism}
\end{equation}
Here, $\alpha$, $C_0$, and $\beta$ are the three free parameters representing the low-mass slope, characteristic halo mass, and redshift evolution, respectively. The high-mass slope is set to zero to reproduce the metallicity saturation observed at $z \sim 0$, and the normalization is set to one, as the maximum distribution fraction is 100\%. We use a simple parameterization for redshift scaling that effectively shifts the characteristic halo mass as a power law with redshift, as the mass coverage of observed MZRs at high redshifts is not wide enough to constrain both the low- and high-mass slopes.
Of note, only the observational constraints for metallicity measurements in the star-forming regions and the CGM are used to constrain the  well-mixed ISM model, while observed metallicities in the \HI\ regions are not used.
\subsection{Multi-phase Metallicity Model}
\label{sec:realistic_model}
We also explore a more realistic and flexible approach where we separate the metallicity in molecular and atomic gas. Star formation occurs in the molecular gas, while atomic gas surrounds the star-forming regions. In this model, we define two ISM distribution fractions, \fHH\ and \fHI, representing the fractions of metals that are distributed to the molecular and atomic regions, respectively. We track the evolution of metal mass in these two regions and derive metallicities by dividing by the respective gas masses. Gas accretes onto the \HI\ regions from the CGM or IGM before accreting onto the \HH\ regions from the \HI\ regions. 

Metallicity evolution in the two regions is expressed as follows:
\begin{eqnarray}
    \dot M_{Z,\mathrm{H2}}
    &=& \{yf_\mathrm{H2}
    - Z_\mathrm{H2}(1-R)\}\mathrm{SFR}\nonumber\\
    &+& Z_\mathrm{HI}\dot M_\mathrm{HI\rightarrow H2}
    + Z_\mathrm{H2,merger}\dot M_\mathrm{H2,merger},
\end{eqnarray}
and
\begin{eqnarray}
    \dot M_{Z,\mathrm{HI}}
    &=& yf_\mathrm{HI}
    - Z_\mathrm{HI}\dot M_\mathrm{HI\rightarrow H2}\nonumber\\
    &+& Z_\mathrm{inf}\dot M_\mathrm{inf\rightarrow HI}
    + Z_\mathrm{HI,merger}\dot M_\mathrm{HI,merger}.
\end{eqnarray}

We parameterize \fHH\ and \fHI\ as functions of halo mass and redshift:
\begin{equation}
    f_\mathrm{H2} = \frac{1}{1+(M_\mathrm{h}/C_0)^{\alpha_0}(1+z)^{\beta_0}},
    \label{eq:fh2}
\end{equation}
\begin{equation}
    f_\mathrm{HI} = \frac{1-f_\mathrm{H2}}{1+(M_\mathrm{h}/C_1)^{\alpha_1}(1+z)^{\beta_1}},
    \label{eq:fh1}
\end{equation}
We adopt the same functional form for \fHH\ as for \fISM\ in the well-mixed ISM model, as they are fitted to the same observational data. For \fHI, we use a similar functional form to \fHH\ because the observational data do not provide enough information to justify a different shape. The term $(1-f_\mathrm{H2})$ ensures that \fHI\ does not exceed unity. This results in six free parameters: $\alpha_0,\ \alpha_1,\ \beta_0,\ \beta_1,\ C_0$, and $C_1$. The observational constraints in the star-forming regions, the CGM, and the \HI\ regions are used in the multi-phase metallicity model.
\subsection{Model Fitting and Posteriors}
\label{sec:modelfit}
Any choice of parameters for \fISM\ determines the evolution of the average ISM and CGM metallicities (
$\langle Z_\mathrm{ISM}(M_\mathrm{h}, z)\rangle$ 
and $\langle Z_\mathrm{CGM}(M_\mathrm{h}, z)\rangle$)
for the well-mixed ISM model.  Similarly, any parameter choice for \fHH\ and \fHI, determine
$\langle Z_\mathrm{H2}(M_\mathrm{h}, z)\rangle$,
$\langle Z_\mathrm{HI}(M_\mathrm{h}, z)\rangle$,
and $\langle Z_\mathrm{CGM}(M_\mathrm{h}, z)\rangle$,
for the multi-phase metallicity model. To compare with observational data, we convert these functions to functions of stellar mass, using the halo mass function and the observed stellar mass probability distribution function (i.e., the probability of having a stellar mass at a given halo mass), both obtained from the \UM\ catalog. For \HI\ metallicity, since observations use velocity width as a proxy for mass, we convert the \HI\ metallicity (as a function of halo mass) to a function of velocity width.
We assume a log-normal distribution with 0.2 dex scatter for velocity width at a given halo maximum circular velocity, based on the results of \cite{Bird15} who have investigated the relation between DLA's velocity width and halo virial velocity using hydrodynamical simulations.
We compare the predicted metallicities after conversion (
$\langle Z_\mathrm{ISM}(M_*, z)\rangle$ 
and $\langle Z_\mathrm{CGM}(M_*, z)\rangle$ 
for the well-mixed ISM model, or
$\langle Z_\mathrm{H2}(M_*, z)\rangle$,
$\langle Z_\mathrm{HI}(\Delta V, z)\rangle$,
and $\langle Z_\mathrm{CGM}(M_*, z)\rangle$,
for the multi-phase metallicity model) with observational data to calculate the likelihood (i.e., $\exp(-0.5\chi^2)$) that the chosen parameter set matches the observed data. We use a Markov Chain Monte Carlo (MCMC) algorithm \citep{Foreman13} to explore the parameter posterior distribution that is consistent with the observations.

\section{Observational Data}
\label{sec:data}
\begin{table*}[ht]
    \centering
    \begin{tabular}{clcc}
    \hline\hline
        Galaxy Component & Reference & Redshift & Corrected \\\hline
        \multirow{4}{*}{Star-forming regions}
        & \cite{Curti20} & $z\sim0$ & Yes\\
        & \cite{Sanders21} & $z\sim2.3$ \& 3.3 & No \\
        & \cite{Nakajima23} & $4<z<6$ & No \\
        & \cite{Curti24} & $4<z<6$ & No \\\hline
        \HI\ gas & \cite{Moller13} & $0.1<z<5$ & No \\
        CGM & \cite{Prochaska17} & $z\sim0$ & No \\\hline
    \end{tabular}
    \caption{Summary of observational constrains on the mass metallicity relation.}
    \label{tab:obs}
\end{table*}
\subsection{Metallicity Measurements in Star-forming Regions}
\label{sec:obs_h2}
For constraints on metallicity in the \HH\ regions, we use the mass-metallicity relations (MZRs) from emission lines of galaxies at $z\sim$ 0--5.
These metallicities are measured using nebular emission lines that are emitted from star-forming (i.e., \HII) regions. 
Given that the \HII\ and \HH\ regions are physically close to each other, we assume that the metallicity in the \HH\ regions is the same as that in the \HII\ regions.
\subsubsection{Local Universe}
Various studies have investigated the MZR of galaxies in the local Universe, employing different methods, including the direct $T_\mathrm{e}$ method and strong-line calibrations. The direct method, based on electron temperature measurements, is widely regarded as more reliable. However, its reliance on detecting faint auroral lines such as \OIII$\lambda$4363 may introduce a bias towards low-metallicity (and thus low-mass) galaxies within MZR samples. In contrast, strong-line calibrations can be applied for a broader range of galaxies, thus reducing sample bias. Despite the larger systematics in individual estimations compared to the direct method, it is crucial to rely on observational data reflecting the true average for constraining the average metallicity of galaxies. Therefore, employing the MZR with strong-line calibrations is more suitable for our purpose.

We use the MZR of \cite{Curti20}, which determined the metallicities for individual SDSS galaxies at $z\sim 0$ and reported the metallicity distribution (i.e., median and scatter) in bins of stellar mass. The metallicities are determined using the strong-line calibrations presented in \cite{Curti17}. The total number of galaxies in this sample is 151862.
{The slope and saturation metallicity of the \cite{Curti20} MZR are consistent with other MZR determinations using the $T_e$ method \citep[e.g.,][]{AM13}, while the normalization is significantly lower compared to the results from photoionization models \citep[e.g.,][]{Tremonti04}.}

The observational samples still have some biases towards high sSFR galaxies in the low-mass regime, due to a minimum SFR selection in \citealt{Curti20}. To account for the biases, we correct the \cite{Curti20} data to estimate the true average metallicities of star-forming galaxies in the local universe. We can do this by assuming that unobserved galaxies lie on the mass-metallicity-SFR relation (i.e., FMR), in which galaxy metallicity is only a function of the parameter $\mu \equiv \log M_* - 0.66 \log\mathrm{SFR}$. Galaxies with SFRs below the cut in \cite{Curti20} will have higher $\mu$ than galaxies above the cut, so no extrapolation of the shape of $Z(\mu)$ is needed. 

In practice, we calculate the average and scatter of sSFR using the \UM\ catalog at $z\sim0$ in each stellar mass bin. We assume a log-normal distribution for sSFRs. As above, we also estimate the metallicity distribution based on the sSFR distribution assuming the mass-metallicity-SFR relation defined by \cite{Curti20} (for total SFR). By integrating the sSFR and metallicity distributions in the range of sSFR values that are not covered by \cite{Curti20}, we determine the fraction and average metallicity values for galaxies below the SFR cut in \cite{Curti20}. Finally, to derive the average metallicity of all star-forming galaxies in bins of stellar mass, we perform a weighted summation using the fraction above, by combining the estimated metallicity values for unobserved galaxies with the metallicity values for observed galaxies. 

We estimate the uncertainties of the average metallicity, as \cite{Curti20} do not provide this information, but rather scatter of the metallicity distribution. The typical systematic uncertainty of calibrations reported in \cite{Curti20} is 0.12 dex, while the typical measurement error of SDSS galaxies is 0.02 dex \citep{AM13}. The mean sum of squares of these two is 0.12 dex. Hence, we adopt an uncertainty of 0.12 dex for every stellar mass bin.
\subsubsection{High Redshifts}
We focus our modeling of metallicity evolution within the redshift range of $0 < z < 5$ due to limited gas measurements beyond this range. We utilize the MZR reported at $z \sim 2.3$ and $z \sim 3.3$ from \cite{Sanders21}, as well as at $z \sim 5$ from \cite{Nakajima23}. We apply no correction for biases in the high-redshift samples, as the galaxies show consistency with the main sequence galaxies at their corresponding redshifts.

\cite{Sanders21} have determined metallicities for stacked spectra of galaxies from the MOSDEF survey \citep{Kriek15}, which were binned by redshift and stellar mass. The total number of galaxies in the sample is 265 at $z\sim2.3$ and 130 at $z\sim3.3$. To determine metallicity, \cite{Sanders21} employ strong-line metallicity calibrations derived in \cite{Bian18}. We adopt the root mean square of the measurement uncertainty and the calibration scatter as the uncertainty of the metallicity measurements. The typical uncertainty is 0.11 dex.

\cite{Nakajima23} have derived metallicities for 135 galaxies at redshifts $z=$ 4--10 using JWST/NIRSpec early release data. The direct method is employed to galaxies with \OIII$\lambda$4363 detection, while strong-line metallicity calibrations of \cite{Nakajima22b} are applied for galaxies without \OIII$\lambda$4363 detection. 
We also include metallicity measurements for 62 low-mass galaxies from JWST Advanced Deep Extragalactic Survey (JADES; ref) reported by \cite{Curti24}. The determination of gas-phase metallicity utilizes a modified version of the \cite{Curti20} calibrations, incorporating new diagnostic methods suggested by \cite{Laseter24}. We use 104 galaxies in total within the redshift range of $z=$ 4--6 (median at $z \sim 5$), 62 from \cite{Nakajima23} and 34 from \cite{Curti24}. We adopt the median and its uncertainties within three stellar mass bins for the individual metallicity measurements. These uncertainties combine the measurement uncertainties and calibration uncertainties. The typical uncertainty is 0.23 dex.

\subsection{Metallicity Measurements in Atomic Gas}
\label{sec:obs_hi}
DLAs are considered to trace dense \HI\ gas within galaxies. We rely on metallicity measurements of DLAs across a range of redshifts ($z =$ 0--5) compiled in \cite{Moller13}. These measurements are derived from the analysis of metal absorption lines, including those of zinc (Zn), sulfur (S), silicon (Si), and iron (Fe), and are reported as [X/H]. An offset of 0.3 dex has been applied to [Fe/H] measurements to correct for $\alpha$-enhancement \citep{Rafelski12}. We compare the oxygen abundance predicted by the model with these observational data, assuming solar abundance {of \cite{Asplund09}}.

The velocity widths of low-ionization lines are used as a proxy for the halo mass. \cite{Bird15} have demonstrated using cosmological hydrodynamic simulations that the relation between the velocity width of DLAs and their virial velocity is order of unity, despite large scatter. Based on their results, we convert virial velocities into observed velocity widths assuming a log-normal distribution with a scatter of 0.2 dex and no offset (see also Section \ref{sec:modelfit}).

We divide the individual DLA measurements into bins of redshift ($z \sim$ 1, 2, 3, and 4) and velocity widths. Then, we estimate the average metallicities and their uncertainties in each bin, using the bootstrap estimation method. More specifically, we derive the average metal number density $\langle n_\mathrm{X} \rangle$ and hydrogen density $\langle n_\mathrm{H} \rangle$, which are then combined to yield the average metallicity $\langle \mathrm{X}/\mathrm{H}\rangle=\langle n_\mathrm{X} \rangle / \langle n_\mathrm{H} \rangle$.

It should be noticed that these data do not account for the effects of dust depletion. Of the DLA measurements used in this study, $\sim$70\% (76 out of 110) rely on Zn and S, which are considered non-refractory elements with minimal dust depletion. The remaining measurements involve Si and Fe, which exhibit significant depletion, especially at high metallicity. \cite{DeCia18} provide the amount of dust corrections for some DLAs used in this work. We confirm that the average metallicities in each velocity width and redshift bin, before and after the dust correction, agree within the uncertainties, except for the highest velocity and redshift bin. Since the sample size for dust depletion correction is small in the highest redshift bin (7), we use the uncorrected values reported in \cite{Moller13}.
\subsection{Metallicity Measurements in the CGM}
\label{sec:obs_cgm}
We utilize metallicity measurements in the CGM at $z\sim0.2$, obtained from quasar spectra in the COS-Halos survey \citep{Tumlinson11,Tumlinson13}, as reported in \cite{Prochaska17}.  We divide the sample into three stellar mass (from \citealt{Werk12}) bins and apply a similar methodology to that used for DLAs (as described in Section \ref{sec:obs_hi}) to estimate the average metallicity and its uncertainty within each bin.
\section{Results}
\label{sec:result}
\subsection{Fitting Results}\label{sec:fitting}
\begin{figure*}[t]
    \centering
    \includegraphics[width=\linewidth]
        {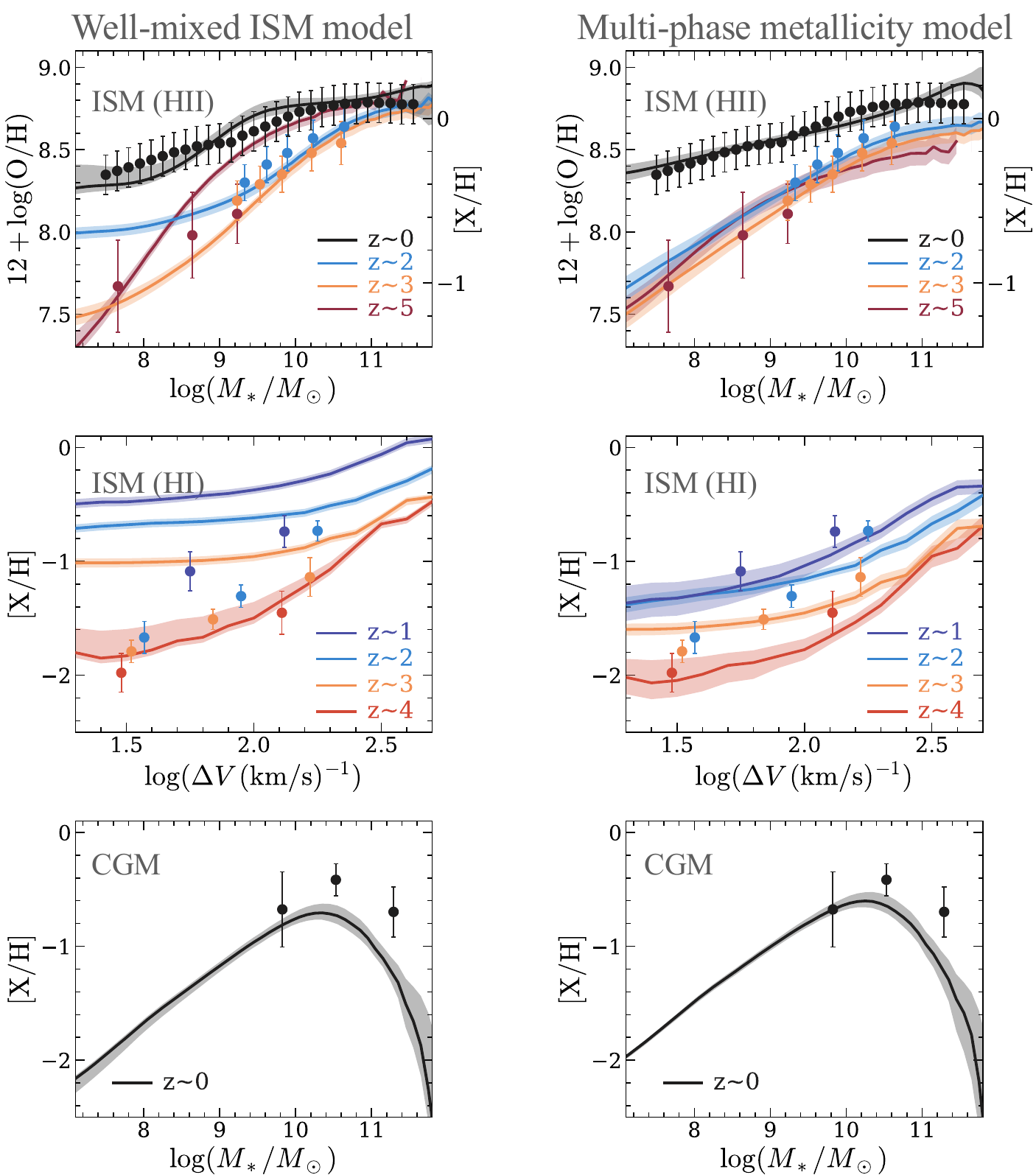}
    \caption{
    {Comparison between the models (curves) and the observed mass-metallicity relations (circles). Left: The well-mixed ISM model in star-forming regions (top), \HI\ regions (middle), and the CGM (bottom). The shaded regions represent 68th percentile of the posterior distributions. Colored circles represent observational data at $z \sim 0$ \citep{Curti20}, $z \sim 2$, $z \sim 3$ \citep{Sanders21}, and $z \sim 5$ \citep{Nakajima23,Curti24} in the top panel; $z \sim 1-4$ \citep{Moller13} in the middle panel; and $z \sim 0$ \citep{Prochaska17} in the bottom panel.}
    Note that the observational data of \HI\ metallicity is not used for the model fitting. Right: Same as the left panel, but for the multi-phase metallicity model, with additional observational constraints of \HI\ metallicity.
    {This model shows better fits to the observational data.}
    }
    \label{fig:mzr}
\end{figure*}
%
%
%
Our model is combined with four models, each making different assumptions about inflow sources (either IGM or CGM; see Section \ref{sec:inflows_cgm}) and mergers (either 100\% of mergers bringing metals to the host galaxy’s ISM or no contribution from mergers; see Section \ref{sec:method_mergers}), for both the well-mixed ISM model and the multi-phase metallicity model. For each model, 
we performed an MCMC sampling with 32 walkers and 1000 steps per walker, exceeding 10 times the autocorrelation time to effectively explore the posterior distribution.
For reviews of the dependencies of the results on the assumptions about inflow sources and mergers, refer to \ref{sec:each_model}. 
Subsequently, we resampled the ISM distribution fractions and metallicity evolution 1000 times by randomly selecting models from the four posterior distributions without weighting. The best-fitting models are determined based on the median values obtained from the resampled models.
The best-fitting parameters and their 68th percentiles of the posterior distributions for the well-mixed ISM model are
$[\log C,\ \alpha,\ \beta] = 
[14.5_{-4.1}^{+0.3},\ -0.11_{-0.03}^{+0.22},\ -1.8_{-0.6}^{+0.6}]$,
and for the multi-phase metallicity model are 
$[\log C_0,\ \alpha_0,\ \beta_0,\ \log C_1,\ \alpha_1,\ \beta_1] =
[13.3_{-0.3}^{+0.4},\ -0.26_{-0.05}^{+0.05},\\ 0.5_{-0.2}^{+0.1},\ 13.2_{-0.8}^{+0.9},\ -0.6_{-0.2}^{+0.2},\ -1.6_{-0.9}^{+0.8}]$,
respectively.

\begin{figure*}[ht]
    \centering
    \includegraphics[width=\linewidth]
        {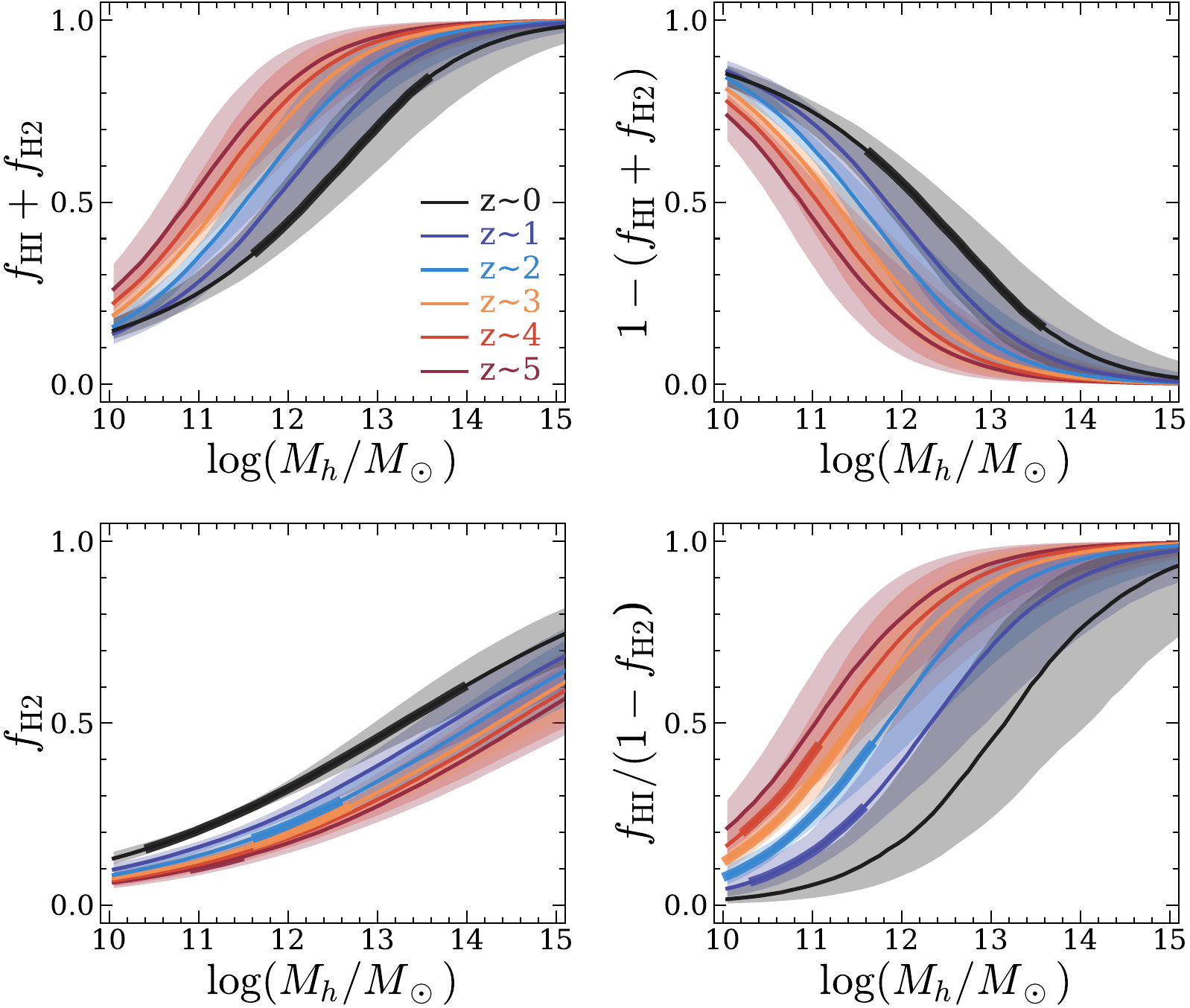}
    \caption{The ISM distribution fractions of the multi-phase metallicity model. Each panel shows the distribution of metals within the multi-phase gas of the galaxy, including the total distribution fraction to the ISM (i.e., either \HH\ or \HI\ regions; upper left), the CGM (upper right), \HH\ regions (lower left), and the fraction to \HI\ among metals that are not distributed to \HH\ (lower left). In all panels, thick lines show the mass ranges covered by the observational data. The shaded regions represent 68th percentile of the posterior distributions.}
    \label{fig:fism_realmix}
\end{figure*}

We show the mass-metallicity relations of the well-mixed ISM model in left panel of Figure \ref{fig:mzr}. The observational data from star-forming regions (top panel) and the CGM (bottom panel) are incorporated as observational constraints, while data from \HI\ regions (middle panel) are not included in the fitting procedure.
The naive reduced $\chi^2$ of the best-fitting model is {4.6 for the number of observational data points including the \HI\ metallicities ({54}). The largest discrepancy occurs at $z\sim5$ in star-forming regions, where the model shows higher metallicity compared to the observed metallicity at higher-mass points, and even surpasses the predicted metallicity of the model at $z\sim3$ for the given mass. The elevated metallicities at $z\sim5$ probably arise from the slight decrease of ISM (specifically \HI) masses from $z\sim4$ to $z\sim5$ in the \NUM\ model. Note that this decreasing trend is unclear in the observed \HI\ cosmic density, due to large uncertainties and scatter in the observational data (see Figure 4 of \citealt{Guo23}). 

In addition, this model fails to reproduce the observed MZR in \HI\ regions at $z\lesssim3$ (noting that the observational data are not used in the model fitting process), as indicated by the left panel of Figure \ref{fig:mzr}. Since the observed metallicity in \HI\ regions is lower than that in star-forming regions at a given mass, the model predicts higher metallicity in \HI\ regions. These discrepancies 
imply that the assumption of metals being homogeneously mixed within the ISM is too simplistic.  This is not just an artifact of our modeling, but can be readily validated from the data directly.  For example, the ISM metallicity of a $10^9$\Msun{} galaxy at $z=2$ is equivalent to $[X/H]\sim -0.5$, but this galaxy mass corresponds to a $\sim 10^{11}$\Msun{} host halo with $\Delta V \sim 100$ km/s, for which typical DLA metallicities are $[X/H] \sim -1.5$.

We show the mass-metallicity relations of the multi-phase metallicity model in right panel of Figure \ref{fig:mzr}. This model shows substantially better agreement with the MZRs for all gas phases, including star-forming regions, \HI\ regions, and the CGM. For the number of observational data points we use (54), the naive reduced $\chi^2$ of the best-fitting model is 0.60, suggesting a reasonable fit.
\subsection{Distribution of Metals between the ISM and CGM}
\label{sec:fism}
%
In Figure \ref{fig:fism_realmix}, we show the fraction of metals distributed to the ISM for the multi-phase metallicity model. 
The top left panel illustrates the total metal distribution fractions to all phases of the ISM, i.e., \fHH\ $+$ \fHI. The top right panel shows the distribution fraction to the CGM, $1-(f_\mathrm{H2}+f_\mathrm{HI})$.
The bottom left panel displays the metal distribution fraction to \HH, i.e., \fHH. The bottom right panel shows the fraction of metals distributed to \HI\ compared to those not distributed to the \HH\ regions, i.e., $f_\mathrm{HI} / (1-f_\mathrm{H2})$. In these panels, the mass range covered by the observational data are shown in thick lines.
The total ISM distribution fraction (\fHH\ $+$ \fHI) gradually increases with halo mass
from $\sim20$\% to $\sim80$\% ($\sim30$\% to $\sim100$\%) in $10^{10}-10^{13} M_\odot$ halos at $z=0$ ($z=5$).}

At $z\sim0$, the \HI\ distribution fractions remain low (\fHI\ $\lesssim 0.2$) across mass, indicating that a higher fraction of metals are distributed to \HH\ rather than \HI. This suggests poor metal mixing between the \HH\ and \HI{} phases, highlighting the importance of modeling the multi-phase gas contents of molecular and atomic gas separately. 

The total distribution fractions exhibit a gradual increase towards high redshift, with the combination of a slight decrease of \fHH\ and significant increase of \fHI\ with increasing redshift.
For the increasing trend of the \HI\ distribution fraction as well as the decreasing trend of the \HI\ distribution fraction, we discuss physical interpretations in Section \ref{sec:fism_evolution}.
{
For interested readers, we show $f_\mathrm{ISM}$ for the well-mixed model in Appendix \ref{sec:fism_wellmix}.
}
%
%
%
\subsection{Systematic Uncertainties}
\label{sec:uncertainties}
Before interpreting the redshift evolution of the ISM distribution fractions (\fHI\ and \fHH), we should note that our model has many systematic uncertainties beyond the posterior distributions, relating to both measurements and modeling.

First, the normalization of the MZR is sensitive to metallicity calibrations \citep{Kewley08,Hirschmann23}. Notably, significant differences arise at low metallicities (12+log(O/H) $<$ 8). When comparing the metallicity estimated from the R23 index of the \cite{Nakajima22b} calibration (utilized in the $z\sim5$ MZR) to that estimated from the \cite{Curti20} calibration (employed at $z\sim0$), a discrepancy of 0.1 dex emerges at low metallicity (12+log(O/H) $<$ 8), while the two are comparable at higher metallicity (12+log(O/H) $>$ 8). On the other hand, utilizing the R23 calibrations of \citep{Sanders24}, calibrated with JWST high redshift galaxies, results in a potentially lower metallicity by 0.1--0.2 dex at low metallicity (12+log(O/H) $<$ 8) compared to the original \cite{Nakajima22b} metallicity. Further investigation is required to confirm how this impacts the median values and \fISM. If the observed metallicity values at $z\sim5$ would be lower (corresponding to the \cite{Sanders24} normalization), the estimated \fHH\ would correspondingly decrease, strengthening the evolution of \fHH.

There also exist systematic uncertainties in molecular gas measurements, which are utilized to constrain the molecular gas mass in the \NUM\ model. The main systematic uncertainty is the CO-to-H$_2$ conversion factor, $\alpha_\mathrm{CO}$. All CO measurements (listed in table 3 of \cite{Walter20}) except for $z\sim0$ measurements of xCOLD GASS, employ a Galactic conversion. However, high-redshift galaxies may possess a higher conversion factor than local galaxies due to their lower metallicity on average, with theoretical and observational support indicating an expected increase in $\alpha_\mathrm{CO}$ in low metallicity gas \citep{Bolatto13}. If $\alpha_\mathrm{CO}$ is indeed higher at high redshifts compared to $z\sim0$, the molecular gas estimates at high redshifts would increase compared to the original values. This would require a higher H$_2$ distribution fraction in our model, resulting in a smaller redshift evolution for \fHH.

Atomic gas masses at high redshifts ($z>1.5$) are derived from DLA measurements. However, DLA measurements typically do not capture low \HI\ column density ($\log N(\mathrm{HI}) > 20.3\ \mathrm{cm^{-2}}$), potentially resulting in an underestimation of the \HI\ density by 10—20\% \citep{Zafar13,Berg19,Peroux20}. Conversely, the contribution of \HI\ gas outside the ISM increases towards high redshifts \citep{Peroux20}, accounting for approximately 10\% of cosmic \HI\ densities at $z\sim5$ \citep{Villaescusa-Navarro18}. While these effects may partially offset each other, the possibility of significant systematic errors cannot be discounted.  
Additionally, DLA measurements constrain cosmic \HI\ density, which is most effective for constraining the \HI\ mass in 10$^{11}$-10$^{12}$ \Msun\ halos, and the \HI\ content in low-mass halos may require further refinement.

The stellar yield is heavily dependent on the adopted IMF and stellar evolution model. While we assume a fixed value for the oxygen yield of 0.016, it varies from $y=0.005$ to 0.033 depending on the IMF and model \citep{Vincenzo16}. As discussed in \cite{Sanders21}, it is conceivable that the oxygen yield increases with redshift due to potential changes in the IMF, including a higher upper mass cutoff and a shallower high-mass slope at high redshifts. If the yield does indeed increase with redshift, the estimated \fHH\ would decrease at higher redshifts to match the observed metallicity, leading to a larger evolution in \fHH.
{We show the impact of the adopted stellar yield on our modeling in \ref{sec:yield}.}

The \UM\ model carries typical systematic uncertainties of 0.2 dex in stellar mass, arising from assumptions in fitting SEDs \citep[see][for a review]{Conroy13}. Typical star formation histories, metallicities, and dust all change toward higher redshifts. For example, higher-redshift galaxies tend to have rising, burstier star formation histories, which may lead to inflated estimates of stellar masses if the same star formation history priors are used as for low-redshift galaxies. Lower actual stellar masses at high redshifts would lead to higher inferred \fHH, and hence lower redshift evolution in \fHH.
\section{Discussion} \label{sec:discussion}
\subsection{Evolution of ISM Distribution Fractions}
\label{sec:fism_evolution}
\begin{figure*}[ht]
    \centering
    \includegraphics[width=\linewidth]
        {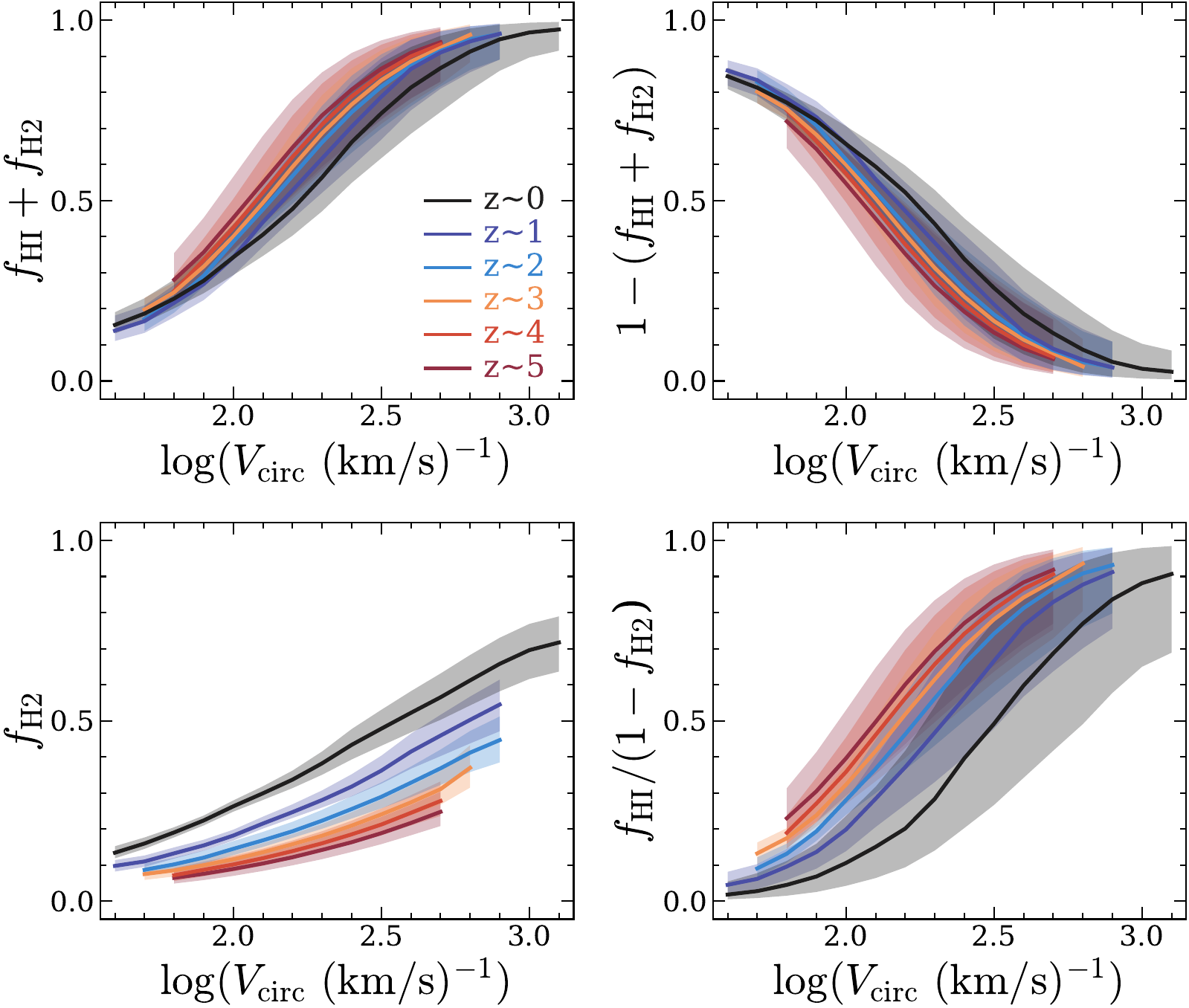}
    \caption{
    Same as Figure \ref{fig:fism_realmix}, but as a function of halo circular velocity.
    }
    \label{fig:fism_vmp}
\end{figure*}
The multi-phase metallicity model indicates an upward trend of the total ISM distribution fraction (\fHI\ $+$ \fHH) with redshift (top left panel of Figure \ref{fig:fism_realmix}). We also evaluate the ISM distribution fractions as a function of halo circular velocity, which represents the gravitational potential well depth of halos (Figure \ref{fig:fism_vmp}). Interestingly, the total ISM distribution fraction (\fHI\ $+$ \fHH) does not evolve with redshift at fixed circular velocity. This indicates a universal trend where the fraction of outflowing gas that reaches the CGM is determined by halo potential well depth only.

In contrast, there is redshift evolution in \fHI\ and \fHH\ both at fixed halo mass and at fixed circular velocity. The \HH\ distribution fraction decreases with redshift by 0.3--0.5 dex from $z=0$ to $z=5$ at \Mh\ $\sim10^{12}$ \Msun, while the \HI\ distribution fraction increases by 0.5--1.2 dex.

We cannot ascribe a cause to this behavior with our current modeling approach.  Nonetheless, we view it as reasonable given at least two physical possibilities:
\begin{itemize}
\item Higher-redshift galaxies have stronger outflows, but also shorter cooling times. For example, \cite{Sugahara19} show that $z \sim 6$ galaxies have higher outflow velocities at a fixed circular velocity. Strong outflows can remove gas and metals from the ISM, which can then quickly recool into the \HI\ phase, leading to a higher fraction of metals distributed to \HI\ vs.\ \HH{}.
\item  Galaxies in the local universe typically exhibit thin stellar disks surrounded by more extended \HI\ gas reservoirs. In contrast, high-redshift galaxies likely possess disordered thick disks because these galaxies are more likely to experience perturbations from increased rates of mergers or inflow from the IGM. Consequently, high-redshift galaxies may have a higher covering fraction of \HI, facilitating easier mixing of gas and metals between \HI\ and star-forming regions.
\end{itemize}

At the same time, given the observational uncertainties in measuring both gas masses and metallicities, as well as their evolution with redshift (Section \ref{sec:uncertainties}), it may also be plausible that improved future measurements show different quantitative trends for the evolution of \fHH\ vs.\ \fHI\ with redshift.
\subsection{Evolution of CGM Metallicity}
\label{sec:fcgm_evolution}
\begin{figure}[ht]
    \centering
    \includegraphics[width=\linewidth]
        {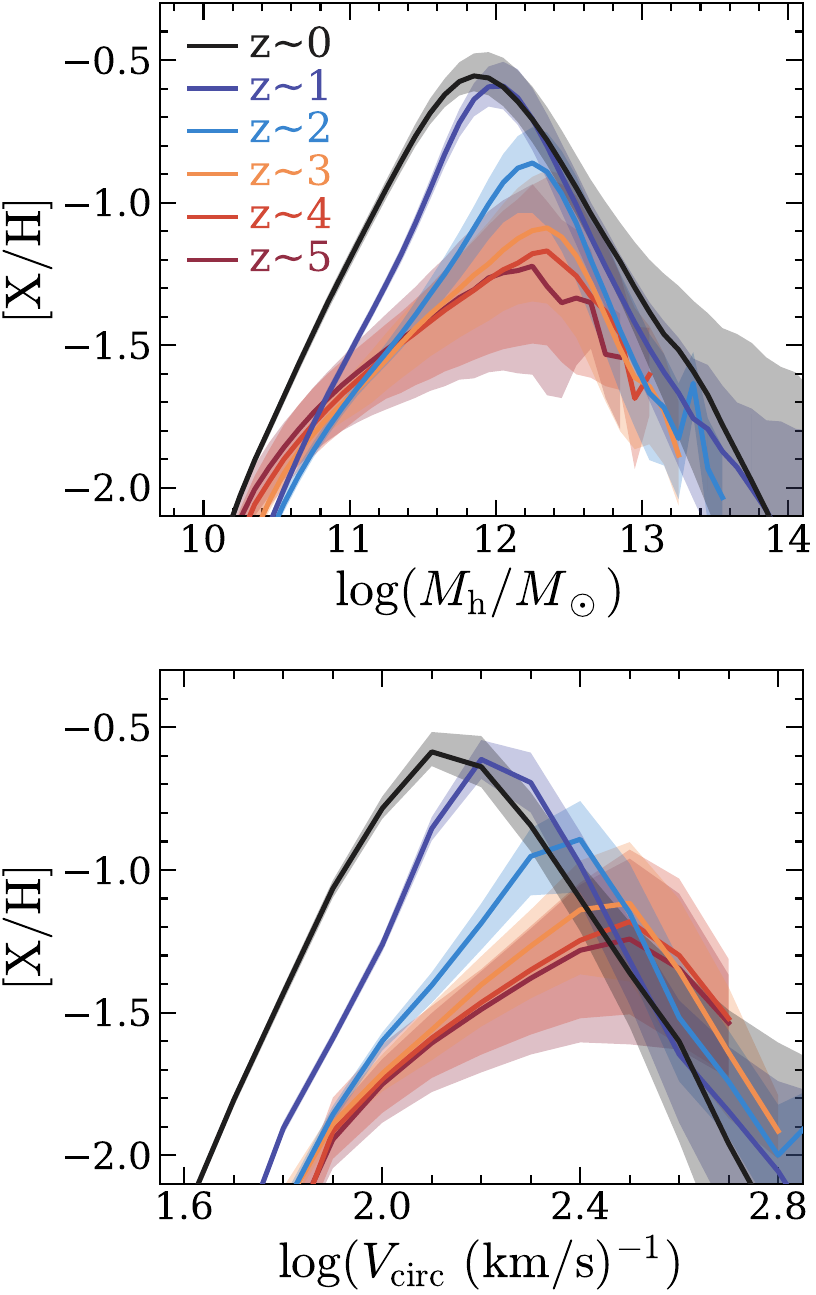}
    \caption{The CGM metallicity evolution of the multi-phase metallicity model as a function of halo mass (top) and those as a function of circular velocity (bottom).  {We expect CGM metallicities to be higher for group and cluster-scale masses than shown here, as we do not include metals ejected from satellites in the host CGM.}}
    \label{fig:cgm_evol}
\end{figure}
Figure \ref{fig:cgm_evol} shows the evolution of the CGM metallicity at $0<z<5$ predicted by the multi-phase metallicity model. It indicates slow enrichment at $z>3$, following rapid enrichment below $z\sim2$. This is attributable to the evolution of the stellar-to-halo mass ratio \citep[see Figure 18 of][]{Behroozi19} combined with increasing the distribution fraction to the CGM towards low redshifts ($1 - (f_\mathrm{HI}+f_\mathrm{H2)}$; right top panel in Figure \ref{fig:fism_realmix}) at fixed halo mass.

Observations of CGM metallicity with Lyman-limit systems indicate a rapid evolution of around 1 dex from $2.3 < z < 3.3$ to $z < 1.0$ (\citealt{Tumlinson17}; see also \citealt{Lehner13} and \citealt{Wotta16}). The magnitude of this observed evolution is consistent with the predictions of our model if the typical host halo mass of Lyman-limit systems changes from \Mh $\sim 10^{11}$ \Msun\ to \Mh\ $\sim 10^{12}$ \Msun.
%
\subsection{Comparison with Mass-Loading Factor}
\label{sec:comparison}
{
Our results suggest that mass-dependent outflows play a key role in shaping the MZR, consistent with previous chemical evolution models \citep[e.g.,][]{Finlator08,Peeples11,Sanders21,Tortora22}.
These studies typically incorporate the mass-loading factor or metal-loading factor, both of which exhibit an inverse correlation with stellar mass and circular velocity. This behavior aligns qualitatively with our findings for $1 - (f_\mathrm{HI} + f_\mathrm{H2})$, as shown in the upper-right panels of Figures \ref{fig:fism_realmix} and \ref{fig:fism_vmp}. While \fISM\ (or $f_\mathrm{HI} + f_\mathrm{H2}$) and the mass-loading (or metal-loading) factor are conceptually similar, a direct quantitative comparison between these parameters is limited due to differences in their definitions.}

{Specifically, \fISM\ accounts for the direct loss of supernova yields, while the mass-loading factor includes the loss of metals from the ISM along with galactic winds. Additionally, the mass-loading factor encompasses only the instantaneous ejection of gas.  Depending on the feedback model, this ejected gas may return to the ISM at a later stage, especially for low-mass galaxies with short cooling times. In contrast, \fISM\ focuses on the total fraction of metals, including those that are directly deposited and those that eventually return to the ISM.}

{
The metals already present in the ISM from previous enrichment episodes and removed by a galactic wind would contribute negatively to \fISM. Although we initially explored an alternative parametrization (i.e., additional normalization parameters to Equations \ref{eq:fism}, \ref{eq:fh2}, and \ref{eq:fh1}) that allowed for negative \fISM\ values, observational data strongly constrain \fISM\ to have positive values. This aligns with the physical picture that a negative \fISM\ would rapidly deplete metals from the ISM, whereas observed galaxy populations show an increasing metal mass over time. 
}
\section{Summary}\label{sec:summary}
We present a model of average metallicities in multiple gas phases, alongside constraints for the distribution of metals produced by stars. The model successfully reproduces the observed evolution of the mass-metallicity-relation in \HII\ regions, DLAs, and the CGM across the redshift range of $0 < z < 5$. Our approach offers a self-consistent picture of the chemical evolution of galaxies over this epoch within the framework of $\Lambda$CDM structure formation. Our results are summarized as follows. 

i) We explore two models for metal mixing in the ISM: one assumes metals are well-mixed within the ISM, and the other treats metals in \HH\ and \HI\ regions separately. The well-mixed model fails to reproduce observed MZRs in star-forming regions at $z\sim5$ and \HI\ regions at $z\lesssim3$, while the multi-phase metallicity model aligns well with observed MZRs in star-forming regions, \HI\ regions, and the CGM. This suggests that the well-mixed assumption is too simplistic to capture the observed chemical evolution of galaxies.

ii) In the multi-phase metallicity model, the fraction of metals distributed to the ISM (i.e., \fHH\ $+$ \fHI) increases strongly with halo mass.  In contrast, the redshift evolution of \fHH\ $+$ \fHI is small, especially when viewed in the halo circular velocity plane.  This suggests that the fraction of outflowing metals distributed to the ISM vs.\ CGM is  determined only by the gravitational potential well depth of the halo.

iii) The distribution fraction to \HH\ decreases with redshift, while the distribution fraction to \HI\ increases. This may result from stronger outflows, higher perturbations due to increased mergers and inflow rates at higher redshifts, and/or observational uncertainties in measuring gas masses and metallicities.
 
%
We thank 
Daichi Kashino, Andrea Pallottini, Daniel Schaerer, Anne Verhamme, Yuichi Harikane, Hidenobu Yajima, Hiroto Yanagisawa, Akinori Matsumoto,  Hiroya Umeda, Minami Nakane, and Tomohiro Yoshida
for useful comments and discussions that improved our manuscript.
This work is supported by JST, the establishment of university fellowships towards the creation of science technology innovation, Grant Number JPMJFS2136. 
Additional support was provided by Overseas Travel Fund for Students (2024) of Astronomical Science Program, The Graduate University for Advanced Studies, SOKENDAI. 
This paper is supported by the World Premier International
Research Center Initiative (WPI Initiative), MEXT, Japan, as
well as the joint research program of the Institute of Cosmic
Ray Research (ICRR), the University of Tokyo. This work is
supported by KAKENHI (20H00180) Grant-in-Aid for Scientific
Research (A) through the Japan Society for the Promotion of
Science.
HG is supported by the National SKA Program of China (grant No. 2020SKA0110100), the CAS Project for Young Scientists in Basic Research (No. YSBR-092) and the science research grants from the China Manned Space Project with NOs. CMS-CSST-2021-A02.
KN acknowledges support from JSPS KAKENHI Grant JP20K22373 and JP24K07102.
The Flatiron Institute is a division of the Simons Foundation.
%
\bibliographystyle{apj}
\bibliography{MAIN.bib}
\appendix
\counterwithin{figure}{section}
\counterwithin{table}{section}
\renewcommand{\thesection}{APPENDIX \Alph{section}}
\renewcommand{\thefigure}{\Alph{section}\arabic{figure}}
\renewcommand{\thetable}{\Alph{section}\arabic{table}}
\section{Corner Plot}\label{sec:corner}
\begin{figure}[ht]
    \centering
    \includegraphics[width=\linewidth]
        {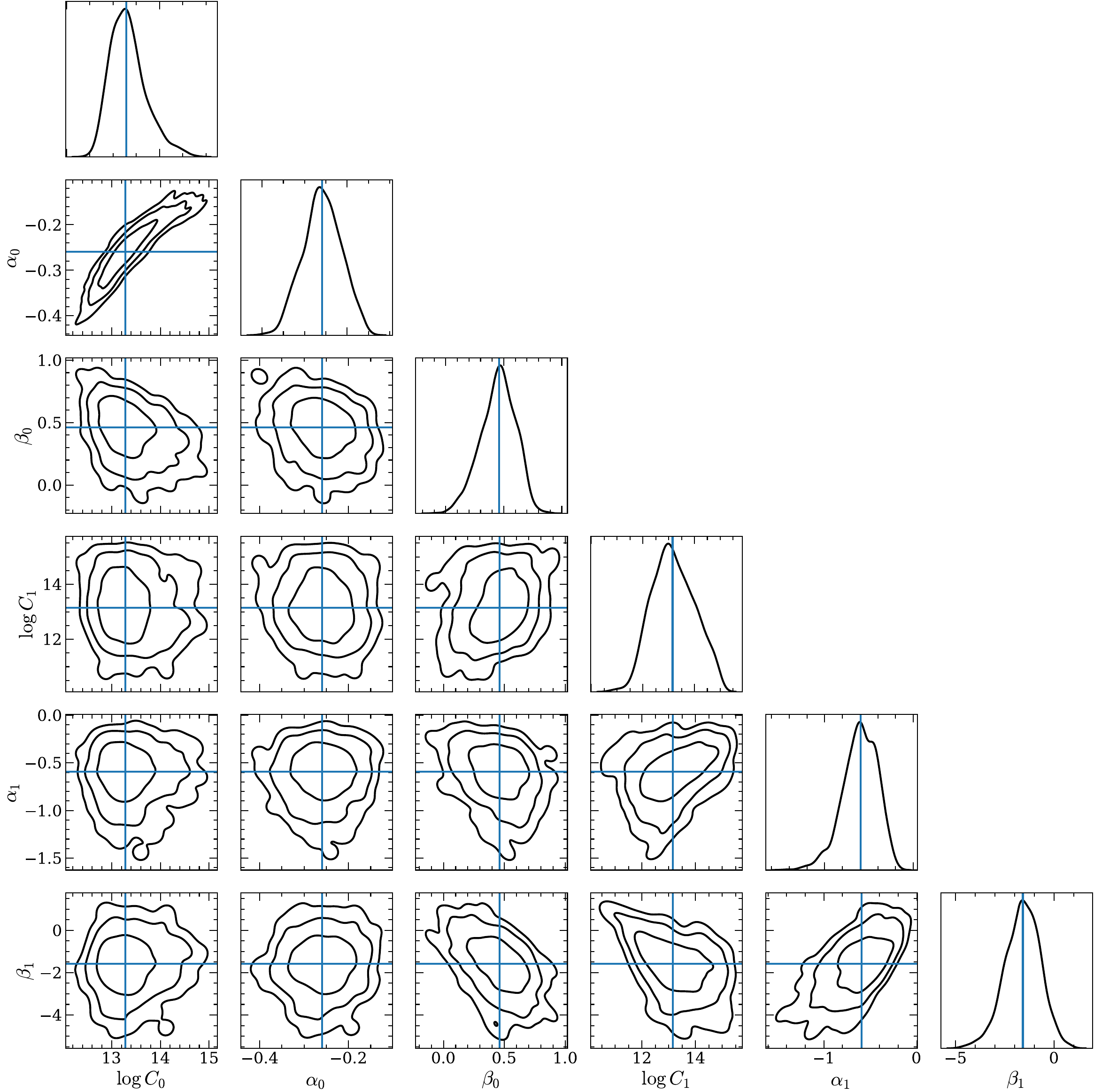}
    \caption{
    Corner plot of the multi-phase ISM model presented in the main text. The blue lines indicate the best-fitting parameters. The contours represent the 1, 2, and 3 sigma levels.
    }
    \label{fig:corner}
\end{figure}

Figure \ref{fig:corner} shows the posterior distributions of the MCMC chains for the multi-phase metallicity model.
\section{Other Fitting Functions}
\begin{figure}[ht]
    \centering
    \includegraphics[width=\linewidth]
        {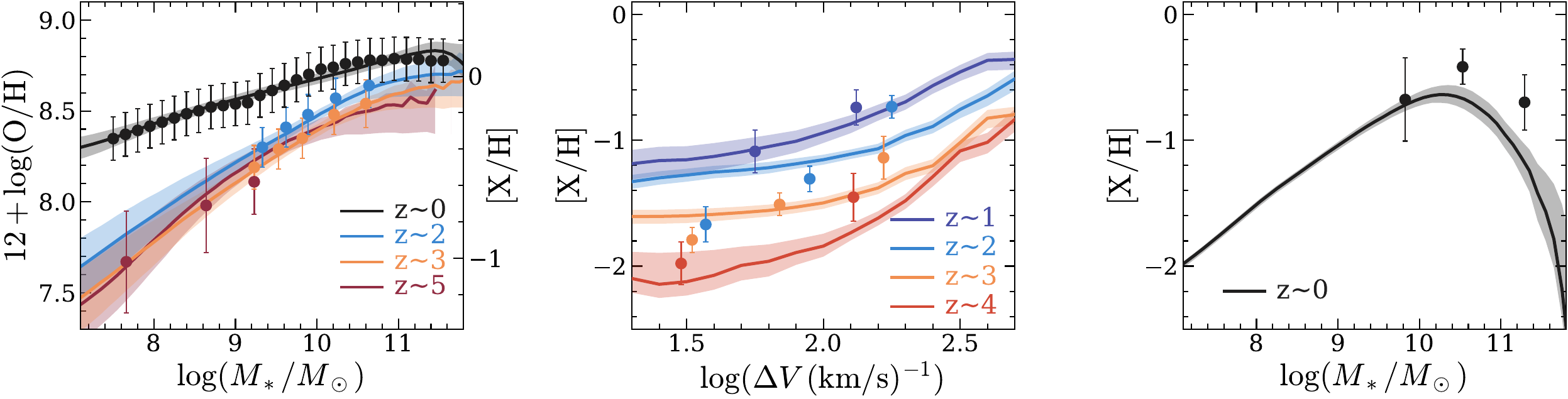}
    \caption{
        {Same as the right panel of Figure \ref{fig:mzr}, but for the other functional form (Equations \ref{eq:c1}--\ref{eq:c4}).}
    }
    \label{fig:mzr_v2}
\end{figure}

\begin{figure}[ht]
    \centering
    \includegraphics[width=\linewidth]
        {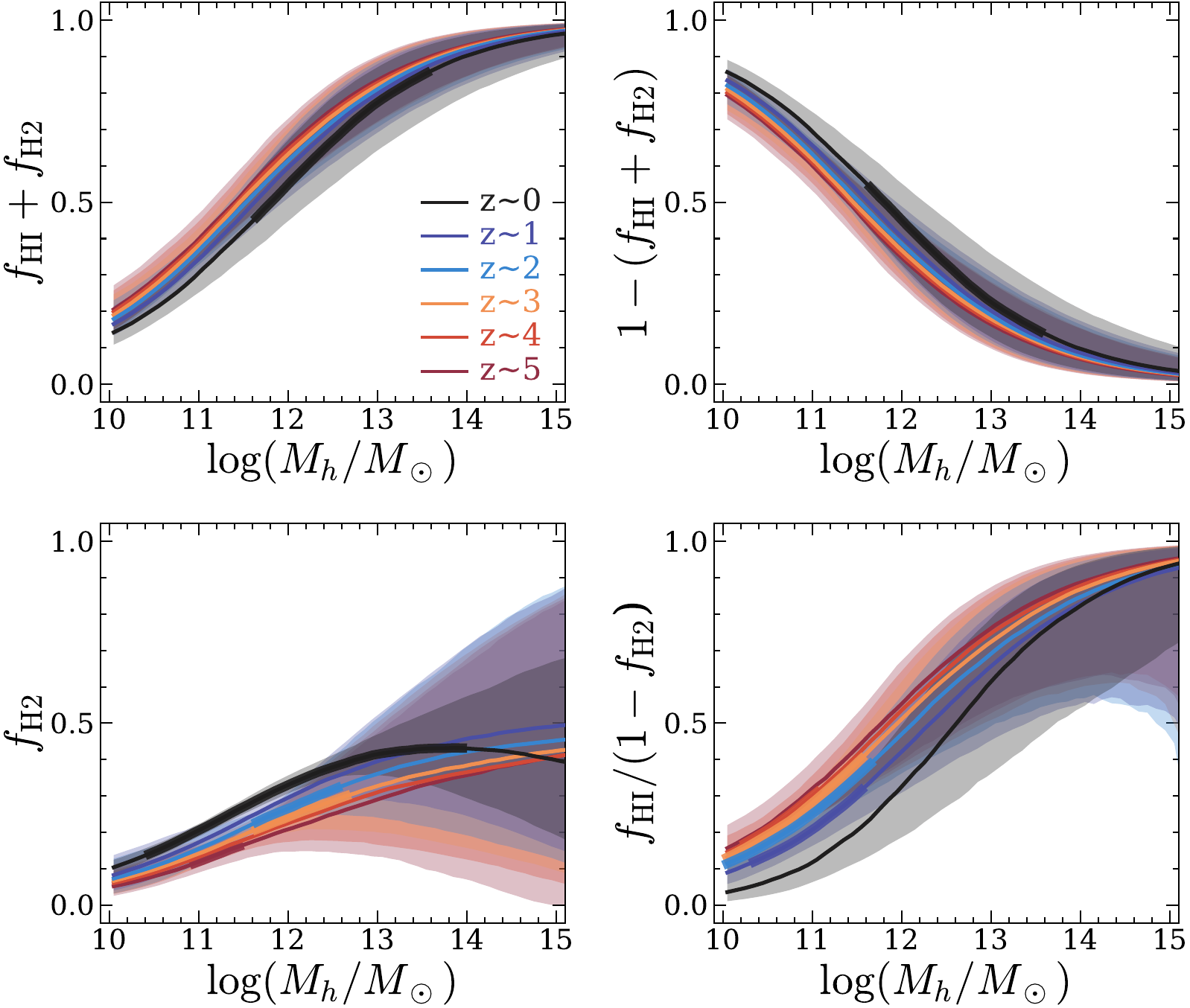}
    \caption{
        Same as the Figure \ref{fig:fism_realmix}, but for the other functional form (Equations \ref{eq:c1}--\ref{eq:c4}).
    }
    \label{fig:fism_v2}
\end{figure}

We investigate alternative fitting forms for \fHH\ and \fHI\ to test the impact of different functional forms on our results, as shown in equations \ref{eq:c1}--\ref{eq:c4}:
\begin{equation}
    f_\mathrm{H2}+f_\mathrm{HI} = \frac{1}{1 + (M_h/C)  ^\alpha \times (1 + z)^\beta}
    \label{eq:c1}
\end{equation}
\begin{equation}
    \frac{f_\mathrm{HI}}{f_\mathrm{H2}+f_\mathrm{HI}} =   \gamma \log M_h + \delta
    \label{eq:c2}
\end{equation}
\begin{equation}
    \gamma = \gamma_0 + \gamma_1 (1-a) + \gamma_2 \ln(1+z)
    \label{eq:c3}
\end{equation}
\begin{equation}
    \delta = \delta_0 + \delta_2 \ln(1+z)
    \label{eq:c4}
\end{equation}
In this section, we consider the total distribution fraction (equation \ref{eq:c1}), which increases with halo mass, and use a linear relation to describe the relative distribution fraction of \HI\ (equation \ref{eq:c2}).
The figures \ref{fig:mzr_v2} shows the best fitting function on the metallicity relations and figure \ref{fig:fism_v2} presents the best fitting functions of the distribution fractions. We find that the functional form does not qualitatively affect our main results: the total distribution fraction increases with halo mass, and its shape is largely independent of redshift.
\section{Effects of Inflow Metallicity and Mergers}
\label{sec:each_model}
\begin{figure}[ht]
    \centering
    \includegraphics[width=\linewidth]
        {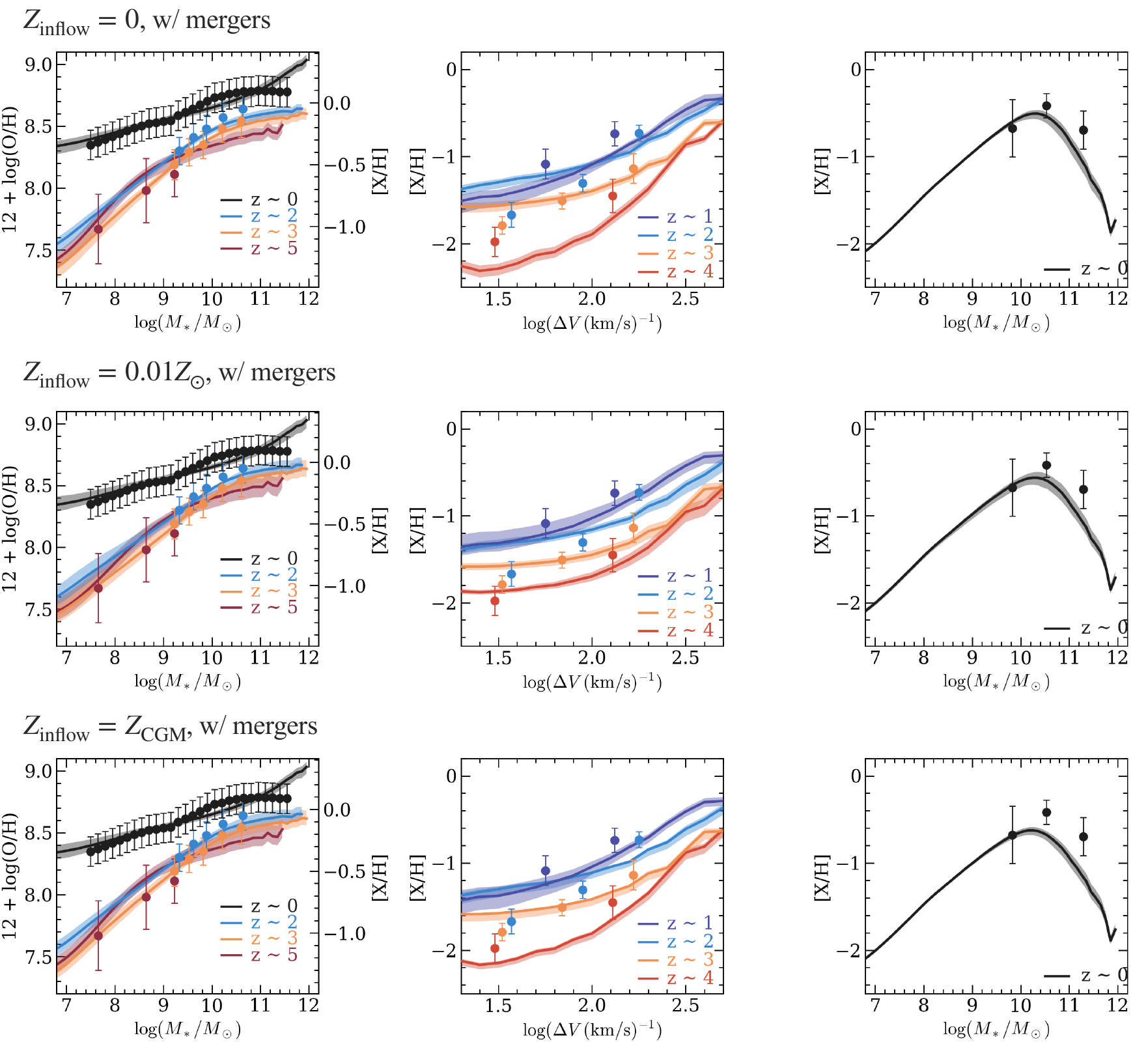}
    \caption{
    Similar to Figure \ref{fig:mzr}, but for the models with mergers, with varying the assumptions of inflow metallicity: $Z_\mathrm{inf}=0$ (top), $Z_\mathrm{inf}=0.01Z_\odot$ (middle), and $Z_\mathrm{inf}=Z_\mathrm{CGM}$ (bottom).    
    }
    \label{fig:mzr_wm}
\end{figure}

\begin{figure}[ht]
    \centering
    \includegraphics[width=\linewidth]
        {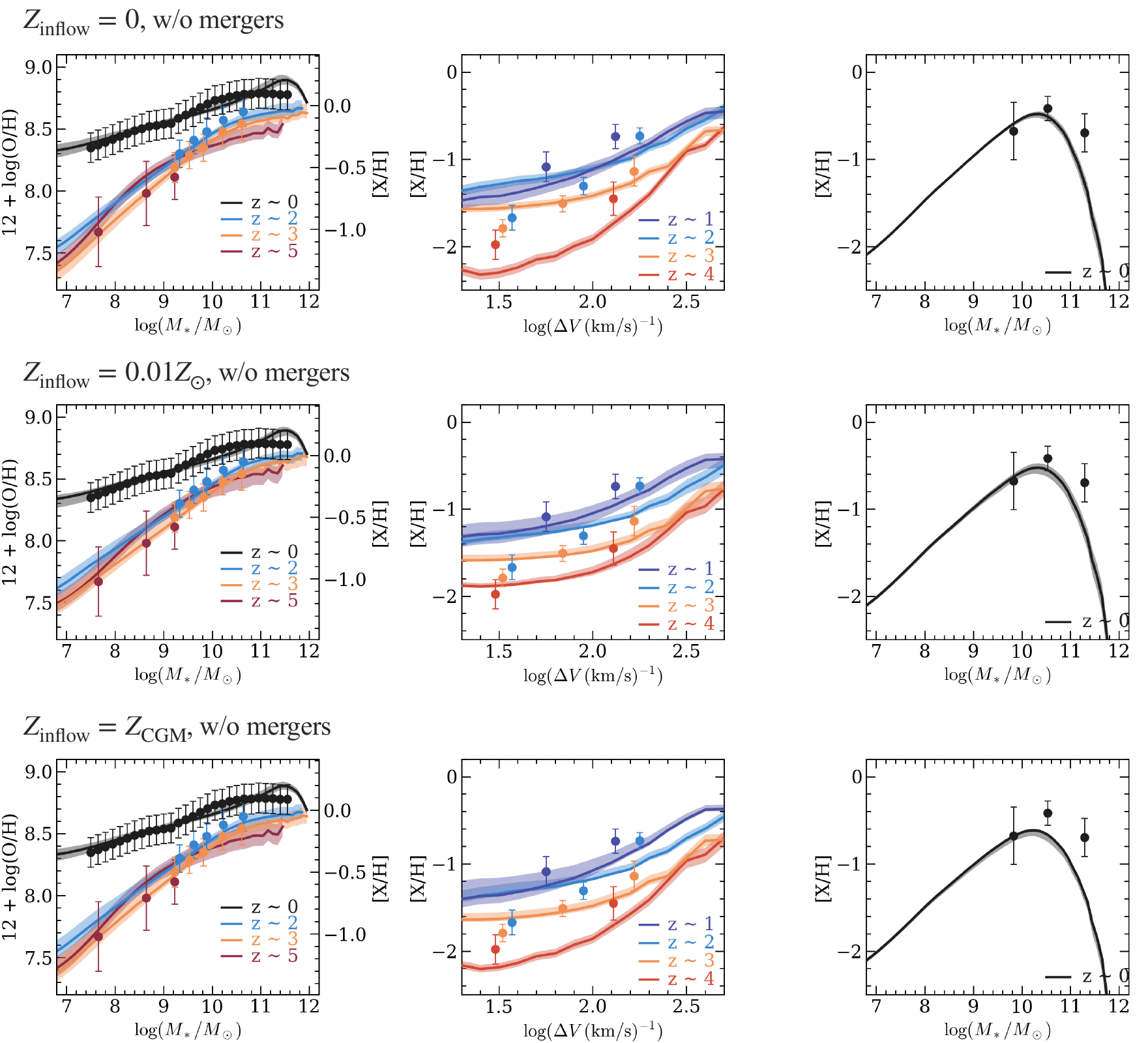}
    \caption{Same as  Figure \ref{fig:mzr_wm}, but for the models without mergers.}
    \label{fig:mzr_wom}
\end{figure}

\begin{figure}[ht]
    \centering
    \includegraphics[width=\linewidth]
        {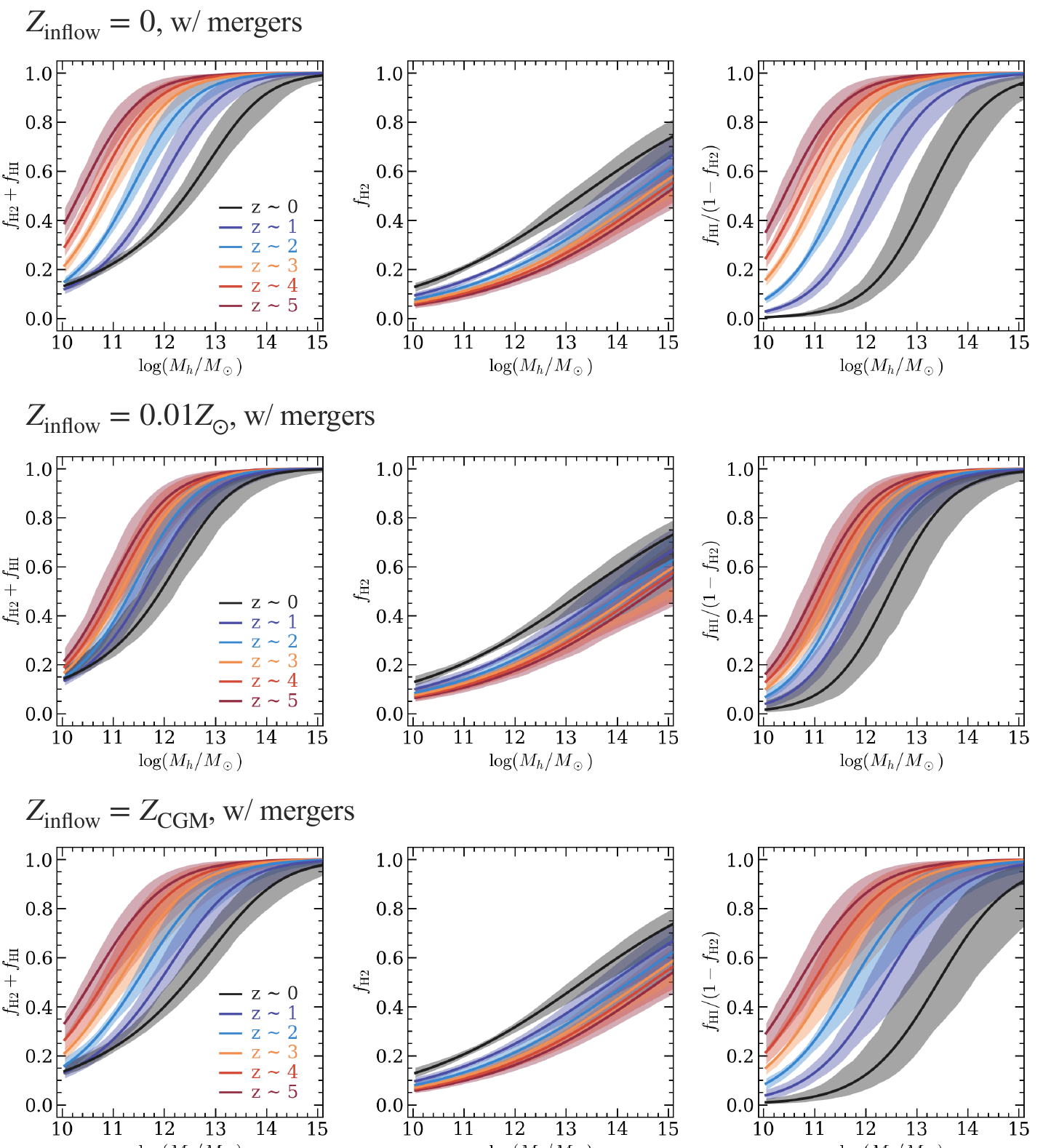}
    \caption{
    Similar to Figure \ref{fig:fism_realmix}, but for the models with mergers, with varying the assumptions of inflow metallicity: $Z_\mathrm{inf}=0$ (top), $Z_\mathrm{inf}=0.01Z_\odot$ (middle), and $Z_\mathrm{inf}=Z_\mathrm{CGM}$ (bottom).
    }
    \label{fig:fism_wm}
\end{figure}

\begin{figure}[ht]
    \centering
    \includegraphics[width=\linewidth]
        {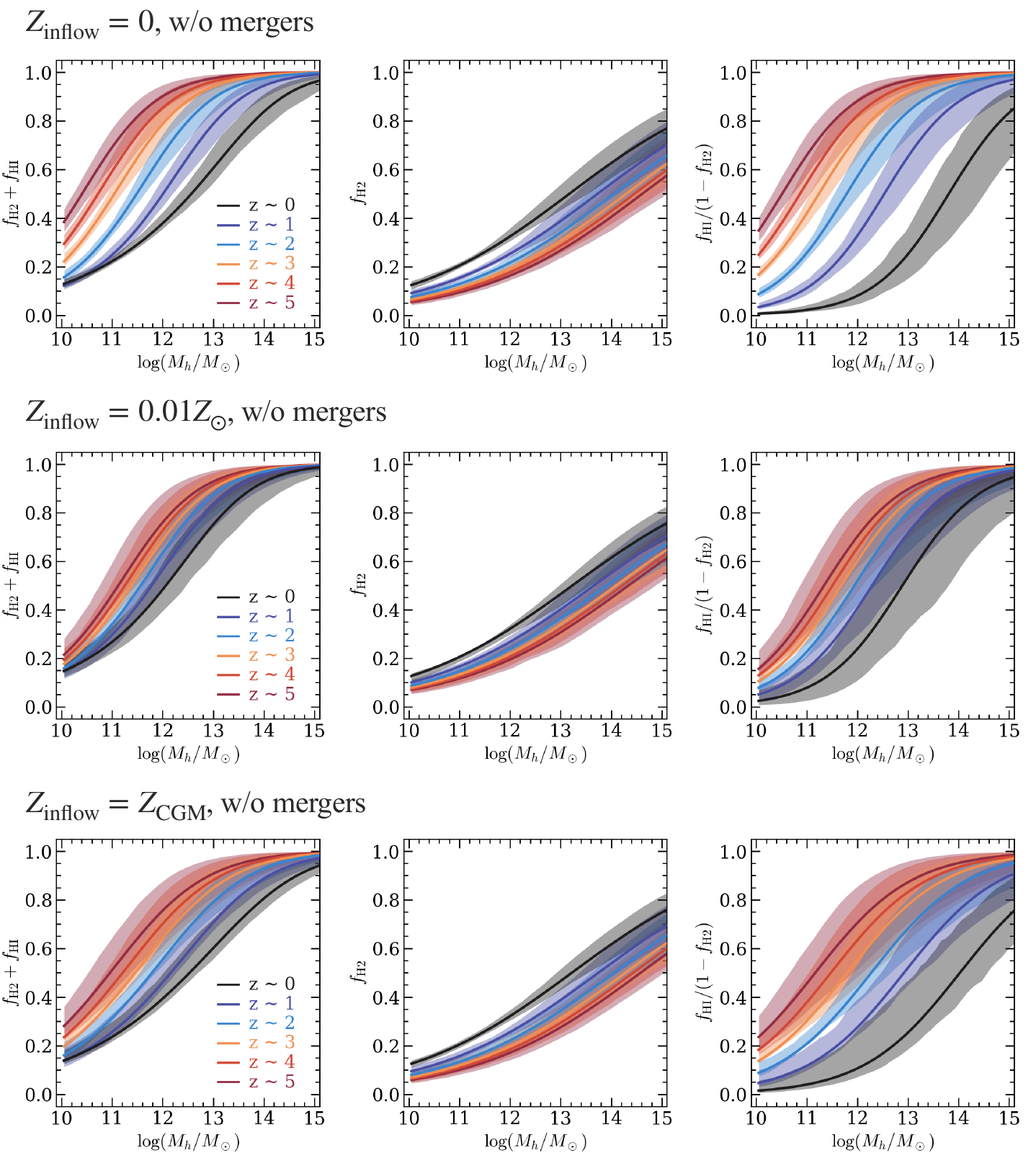}
    \caption{Same as Figure \ref{fig:fism_wm}, but for the models without mergers.}
    \label{fig:fism_wom}
\end{figure}

Figures \ref{fig:mzr_wm}--\ref{fig:fism_wom} show results of models with different assumptions on inflow metallicity and mergers. The best-fitting parameters of each model are shown in Table \ref{tab:best-params}. The metallicity of the inflowing gas influences the metallicity of the \HI\ at low masses at $z\sim4$, while the occurrence of mergers plays a more significant role at higher masses. However, we do not observe any significant differences in the metallicity of \HH\ or the CGM. We find that the functional form does not qualitatively affect the shape of the ISM distribution fractions (Figures \ref{fig:fism_wm} and \ref{fig:fism_wom}).
As described in Section \ref{sec:fitting}, the results presented in the main text are based on a model that randomly samples with equal weighting from four posterior distributions of the models: \(Z_{\text{inf}} = 0.01 Z_{\odot}\) with mergers, \(Z_{\text{inf}} = Z_{\text{CGM}}\) with mergers, \(Z_{\text{inf}} = 0.01 Z_{\odot}\) without mergers, and \(Z_{\text{inf}} = Z_{\text{CGM}}\) without mergers.
\begin{table*}[ht]
    \centering
    \begin{tabular}{cc|cccccc}
    \hline\hline
    Model && $\log C_0$ & $\alpha_0$ & $\beta_0$ & $\log C_1$ & $\alpha_1$ & $\beta_1$ \\
    $Z_\mathrm{inf}$ & mergers &&&&&&\\\hline
    0 & yes & $13.3_{-0.3}^{+0.4}$ & $-0.25_{-0.04}^{+0.05}$ & $0.5_{-0.1}^{+0.1}$ & $13.1_{-0.4}^{+0.6}$ & $-0.8_{-0.2}^{+0.2}$ & $-2.7_{-0.7}^{+0.6}$ \\
    0.01 $Z_\odot$ & yes & $13.4_{-0.3}^{+0.4}$ & $-0.25_{-0.04}^{+0.04}$ & $0.4_{-0.2}^{+0.1}$ & $12.5_{-0.4}^{+0.5}$ & $-0.7_{-0.2}^{+0.2}$ & $-1.4_{-0.9}^{+0.7}$ \\
    $Z_\mathrm{CGM}$ & yes & $13.3_{-0.3}^{+0.4}$ & $-0.25_{-0.04}^{+0.05}$ & $0.5_{-0.1}^{+0.1}$ & $13.4_{-0.6}^{+0.7}$ & $-0.6_{-0.2}^{+0.2}$ & $-1.9_{-0.9}^{+0.7}$ \\
    0 & no & $13.2_{-0.3}^{+0.4}$ & $-0.27_{-0.04}^{+0.05}$ & $0.5_{-0.1}^{+0.1}$ & $13.9_{-0.5}^{+0.6}$ & $-0.6_{-0.1}^{+0.1}$ & $-2.4_{-0.6}^{+0.5}$ \\
    0.01 $Z_\odot$ & no & $13.2_{-0.3}^{+0.4}$ & $-0.27_{-0.05}^{+0.05}$ & $0.4_{-0.2}^{+0.1}$ & $12.9_{-0.6}^{+0.7}$ & $-0.6_{-0.2}^{+0.1}$ & $-1.2_{-0.8}^{+0.7}$ \\
    $Z_\mathrm{CGM}$ & no & $13.2_{-0.3}^{+0.4}$ & $-0.27_{-0.04}^{+0.04}$ & $0.5_{-0.1}^{+0.1}$ & $14.0_{-0.6}^{+0.6}$ & $-0.5_{-0.1}^{+0.1}$ & $-1.7_{-0.8}^{+0.7}$ \\
    \hline
    \end{tabular}
    \caption{Best-fitting parameters of each model.}
    \label{tab:best-params}
\end{table*}
\section{ISM distribution fractions for the well-mixed ISM model}
\label{sec:fism_wellmix}
\begin{figure}[ht]
    \centering
    \includegraphics[width=0.5\linewidth]
        {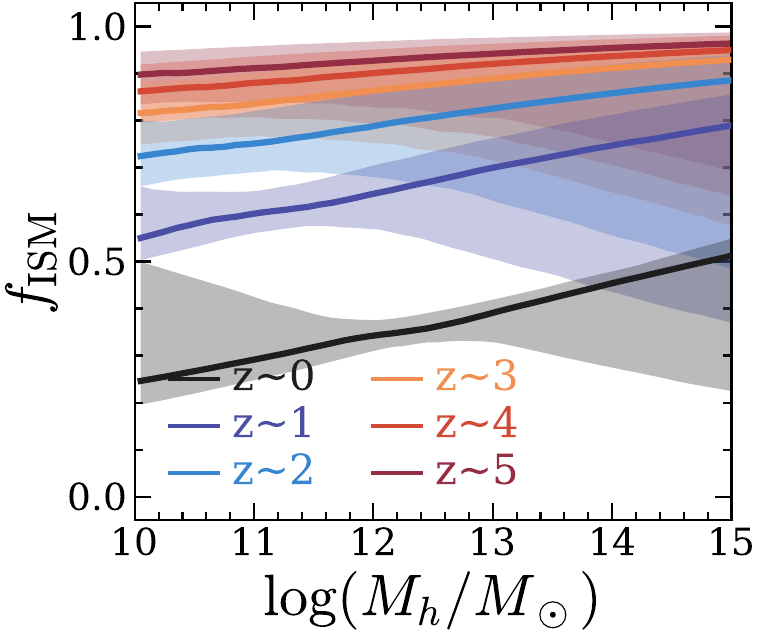}
    \caption{The ISM distribution fraction of the well-mixed ISM model. The shaded regions represent 68th percentile of the posterior distributions.}
    \label{fig:fism_wellmix}
\end{figure}

{
Figure \ref{fig:fism_wellmix} shows the ISM distribution fractions for the well-mixed ISM model. At $z \sim 0$, the ISM distribution fractions increase with halo mass, albeit with the shallow slope  ($\alpha\simeq-0.05$). For redshift evolution, the ISM distribution fractions increase with redshift. Remarkably, at $z\gtrsim1$, the ISM distribution fractions are higher compared to those at $z\sim0$, irrespective of the halo mass, across the mass range of $10<\log M_\mathrm{h}/M_\odot<15$. This suggests that there are some physical mechanism beyond the potential well depth to retain metals in the ISM.}
\section{Effects of Stellar Yield}
\label{sec:yield}
{
The stellar yield is highly sensitive to the adopted IMF and stellar evolution model, as mentioned in section \ref{sec:uncertainties}. While our fiducial value of $y_O=0.016$ is widely used in the literature on the mass–metallicity relation, yields lower than this value are suggested by empirical calibration studies for Milky Way stars \citep[e.g.,][]{Weinberg24}.
}

\begin{figure}[ht]
    \centering
    \includegraphics[width=\linewidth]
        {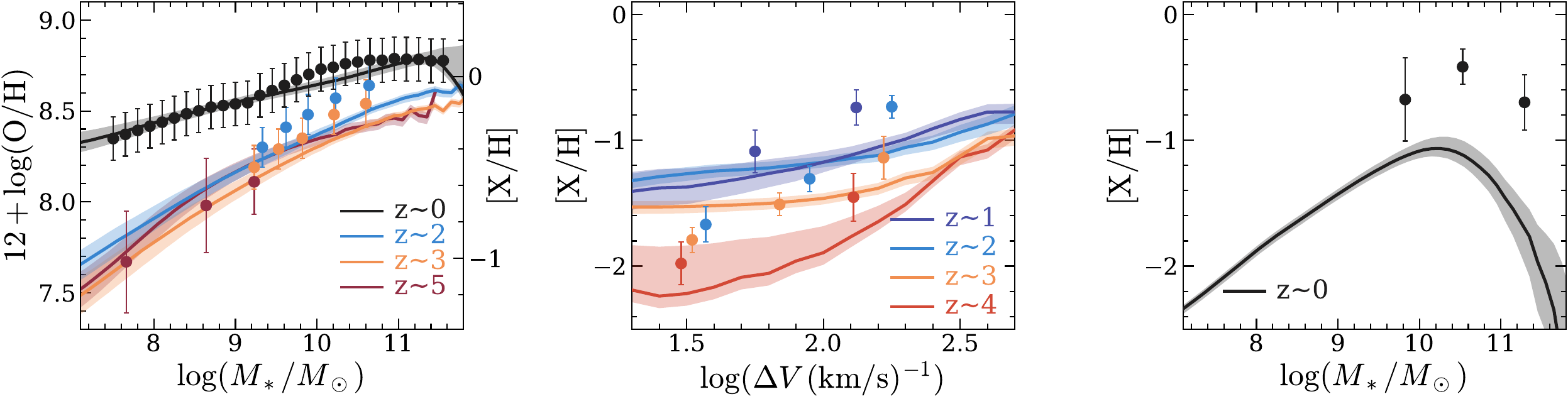}
    \caption{
        {Same as the right panel of Figure \ref{fig:mzr}, but for the lower yield value of $y_O = 0.007$.}
    }
    \label{fig:mzr_lowy}
\end{figure}

\begin{figure}[ht]
    \centering
    \includegraphics[width=\linewidth]
        {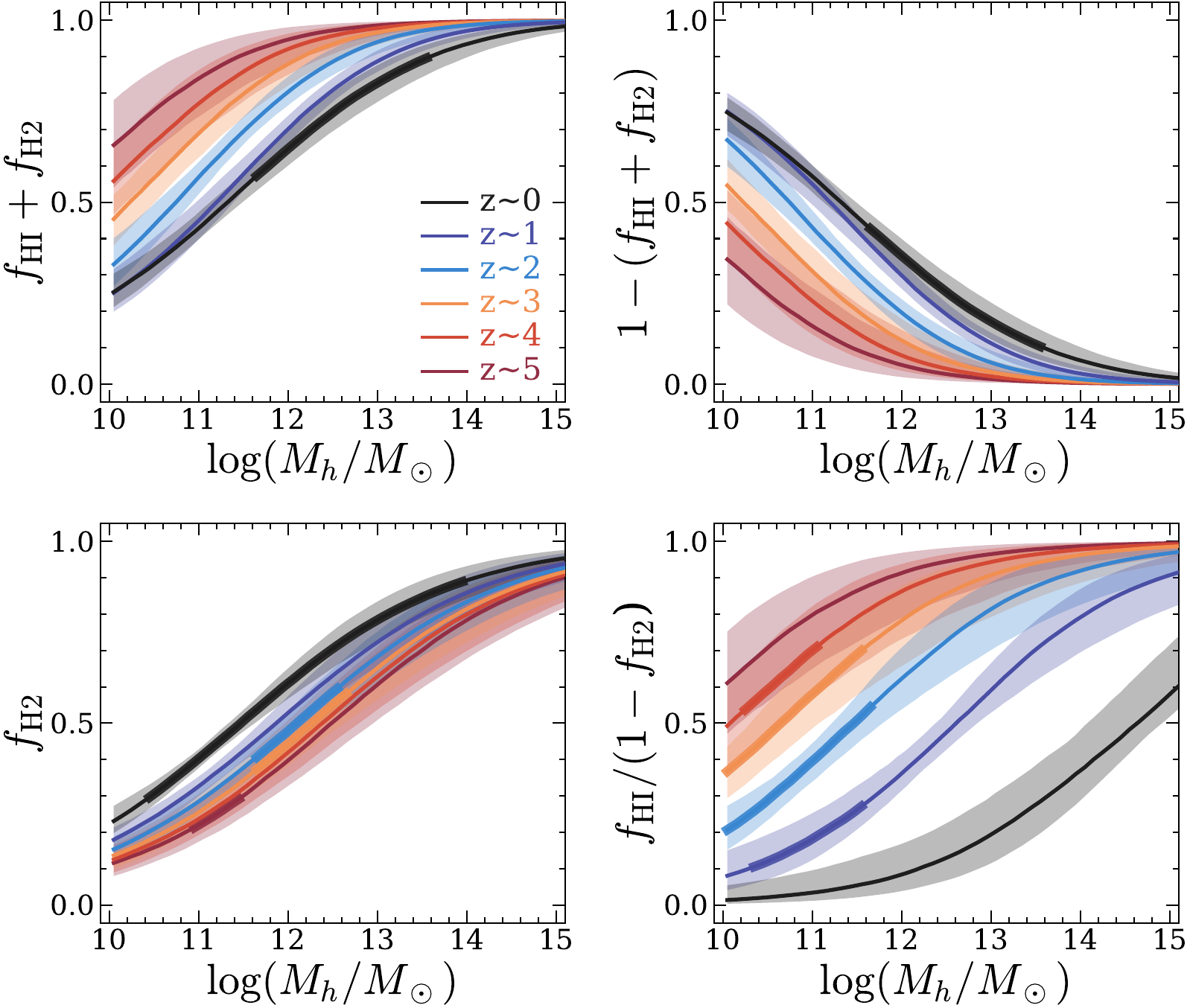}
    \caption{
        {Same as the Figure \ref{fig:fism_realmix}, but for the lower yield value of $y_O = 0.007$.}
    }
    \label{fig:fism_lowy}
\end{figure}

{
To evaluate the impact of the adopted yield on model parameters and fits, we explore a lower yield value of $y_O = 0.007$, which is 2--3 times below the fiducial value. We assume that the yields of other elements are also low, along with $y_O$, when calculating the CGM metallicity. The best-fit models and associated parameters are shown in Figures \ref{fig:mzr_lowy} and \ref{fig:fism_lowy}, respectively.
}

{
With this lower yield, the model provides a poorer fit to observational data, particularly for the CGM metallicity at $z \sim 0$. Since our fiducial model already predicts relatively high $f$-values at large \Mh\ (see Figure \ref{fig:fism_realmix}), the lower yield necessitates even higher $f$-values, which are insufficient to enrich the CGM to observed levels. 
It is important to note that the low $y$ values adopted here are those predicted for oxygen in the Milky Way stars, and it remains uncertain whether other elements exhibit similarly low values. The metallicity measurements in the CGM \citep{Prochaska17} are more strongly constrained by elements such as C, N, and Si than by O. The discrepancy between our model fits and observations may be attributed to non-solar relative abundances within the gas from nucleosynthesis and/or differential dust depletion.
}

{
The model also exhibits a worse fit at $z \sim 2-3$, which could potentially be addressed by adopting a more complex parameterization of the fitting function, where we employ a simple evolutionary formalism to reproduce the metallicity evolution with the fiducial yield (Equations \ref{eq:fh2} and \ref{eq:fh1}).
While a detailed discussion of stellar yields is beyond the scope of this paper, within the framework of this chemical evolution model, assumptions, and formalism, lower stellar yields do not fully account for the observed metallicity relations.
}
\end{document}